\documentclass[pra,floatfix,longbibliography,%
 reprint,
 amsmath,amssymb,
 aps,
]{revtex4-2}

\usepackage{graphicx}
\usepackage{amsmath}
\usepackage{bbm}
\usepackage{braket}
\usepackage{hyperref}
\usepackage{xcolor}

\begin{document}

\title{Excitation Flow, Positivity, and Fisher Information \\
for Open Subsystems of an $N$-Qubit Network}
\author{Tommy Chin}
 \email{wjc5509@psu.edu}
\author{Sarah Shandera}%
 \email{ses47@psu.edu}

\affiliation{Institute for Gravitation and the Cosmos 
and Department of Physics, The Pennsylvania State University, 
University Park, PA 16802, USA}

\date{\today}

\begin{abstract}
An observer of an autonomous quantum system has no privileged initial time, reference state, or subsystem boundary, so the natural framework for the open dynamics is an ensemble of propagators over many subsystems and time windows. We derive such an ensemble analytically for a closed $N$-qubit network with homogeneous all-to-all coupling and a single conserved excitation, giving the propagator of any $K$-qubit subsystem over any time window. To keep the analytical calculation tractable, we exclude excitation flow in sectors of two or more excitations. A single transition amplitude underlies excitation flow, positivity, entanglement entropy, and quantum Fisher information. Positivity and complete positivity reduce to the direction of the excitation flow, so the two coincide for every subsystem size. Both hold for a propagator if and only if it contracts the subsystem state toward a fixed point shared by its dynamical class. Neither the entanglement entropy nor its change over the time window corresponds to this contraction. The ensemble constrains global parameters inaccessible to any single subsystem, such as the size and coupling of the rest of the network. A band of states lies inside the domain of positivity of every single-qubit propagator, yet the reduced state never visits the band. In completely positive divisible dynamics, by contrast, any propagator may act on any state. The quantum Fisher information decomposes into a bounded state contribution and a secularly growing process contribution. The total is minimal when all future propagators are nonpositive and near its maximum when they are positive.
\end{abstract}

\maketitle

\noindent{\small\raggedright\copyright\ 2026 American Physical Society. This is the accepted manuscript of the following article: T.~Chin and S.~Shandera, Phys.\ Rev.\ A \textbf{114}, 052427 (2026), \href{https://doi.org/10.1103/qkr6-slbp}{https://doi.org/10.1103/qkr6-slbp}.\par}

\section{Introduction}\label{sec:intro}

Observers, unlike experimentalists, have no control over the autonomously evolving quantum systems they study. They must infer dynamics and states from partial observations, limited to spatial subsystems, finite time windows, accessible conserved quantities, and finite measurement resolution. Most standard techniques for open dynamics~\cite{Choi1975,Jamiolkowski1972,NielsenChuang2010,Watrous2018,Breuer2002,Rivas2014} presuppose a single well-defined system--environment split and often require control over initial states or coupling strengths. An experimentalist can prepare the system and environment so that their initial correlations are negligible~\cite{HarocheRaimond2006,Blais2021}, and the resulting factorized initial state admits a simpler master equation. A large body of research instead takes the open dynamics to be a quantum channel, a completely positive (CP) and trace-preserving (TP) map, a restriction that removes any dependence on preexisting quantum correlations between the system and environment. Many physical problems admit neither assumption.

\begin{figure}[t!]
    \centering
    \includegraphics[width=0.75\columnwidth]{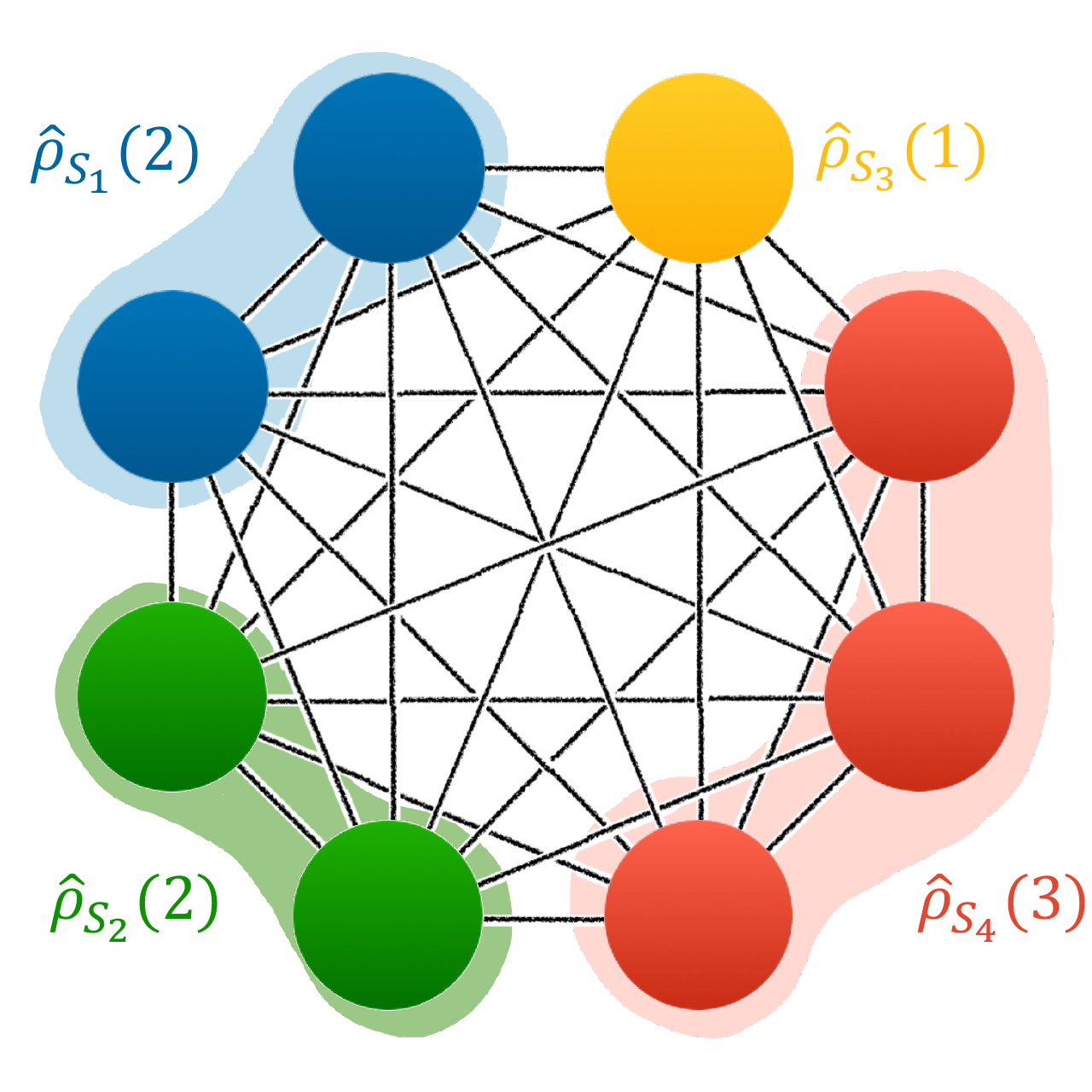}
    \caption{An eight-qubit network with all-to-all interactions. Colored circles denote qubits. Shaded regions indicate possible $K_i$-qubit subsystems described by a density operator $\hat{\rho}_{S_i}(K_i)$.}
    \label{fig:SubSystems}
\end{figure}

The spatial domain and accessible conserved quantities available to an observer establish the subsystem boundary and impose a tensor-product structure on the observable subsystems~\cite{Zanardi_2004}, while the time window sets the temporal boundary. The accessible subsystems and time windows will not in general align with the system's natural temporal or spatial scales, so the resulting dynamics need not be Markovian or CP. Subsystems are generically correlated with each other and with unobserved degrees of freedom, so factorized initial states are the exception rather than the rule. While the process tensor framework~\cite{Pollock2018a,Pollock2018b,Milz2021} provides the general multi-time description of open dynamics and can handle initial quantum correlations, the framework so far describes only a single subsystem.

In this paper, we ask what an observer can infer about an autonomous, finite-dimensional quantum system from observations of correlated subsystems. With no initial time privileged and no reference state available, we propose that the natural framework for such an observer is an ensemble of open dynamics:
\begin{equation}
\bigl\{\Phi_{S_i}(t_1^{(i)},t_2^{(i)})\bigr\},
\end{equation}
where each $S_i$ is an observable subsystem, $[t_1^{(i)},t_2^{(i)}]$ is the observer's time window for $S_i$, and $\Phi_{S_i}$ is the propagator acting on the Hilbert space of $S_i$. This ensemble will generically contain non-CP propagators, and the dynamics of the global network together with the choice of subsystem and time window fix the positivity structure. 

As a concrete example, we study a closed $N$-qubit network evolving under an excitation-conserving, homogeneous, all-to-all Hamiltonian, whose open dynamics are all analytically solvable. The global evolution is unitary, and any subset of $K<N$ qubits constitutes an open subsystem interacting with the remaining $N-K$ qubits. This setting admits closed-form propagators for any subsystem over any time window, making correlations, excitation flow, positivity, and quantum Fisher information all analytically tractable in a unified framework. The single-excitation product state and the conservation law together make all open dynamics phase covariant~\cite{Holevo:1993,Lankinen:2016,Chruscinski2022} and generically nonunital. This analytically solvable example complements prior studies of open-system ensembles from inhomogeneous initial states, where only the CP dynamics for small $N$ was computed~\cite{Prudhoe2024}, as well as numerical studies restricted to $K=1$ subsystems~\cite{Akhouri2025}.

The paper is organized as follows. Section~\ref{sec:nqubit} introduces the closed model. Section~\ref{sec:open-dynamics} derives the propagators for any $K$-qubit subsystem over any time window. Section~\ref{sec:reduced-states} presents the subsystem density operators and their eigenvalue weights. Section~\ref{sec:pcp} establishes when positivity and complete positivity coincide, and obtains the trace distance criterion for any $K$. Section~\ref{sec:geometry} characterizes the ensemble of single-qubit propagators through the Bloch-ball representation, examining the fixed-point distribution and domain of positivity. Section~\ref{sec:subsystem-correlations} computes the entanglement entropy of each subsystem and shows that neither its value nor its change over the time window has a direct relationship to positivity. Section~\ref{sec:complementarity} quantifies sensitivity to global parameters via the quantum Fisher information and decomposes it into state and process contributions. Section~\ref{sec:conclusion} summarizes the results and states open questions for a general observer framework. Appendices~\ref{app:single},~\ref{app:maps}, and~\ref{app:prop} derive the single-excitation amplitudes, the dynamical map, and the propagator. Appendix~\ref{app:choi} computes the Choi spectrum of every propagator.

\section{$N$-Qubit Dynamics}\label{sec:nqubit}

We study a closed network of $N$ qubits evolving under a global unitary. Any subset of $K<N$ qubits constitutes an open subsystem. Tracing out the rest of the network gives the open dynamics of that subsystem, and different choices of subsystem and time window give rise to the ensemble of propagators described in Sec.~\ref{sec:intro}. Figure~\ref{fig:SubSystems} illustrates one such partitioning for $N=8$.

The network evolves under a homogeneous all-to-all XXZ Hamiltonian that conserves the excitation number $q$, the eigenvalue of the operator
\begin{equation}
\hat{Q} = \frac{1}{2}\sum_{i=1}^N \big( \hat{\mathbbm{1}} - \hat{\sigma}^z_i \big),
\end{equation}
where $\hat{\sigma}^z_i$ is the Pauli-$z$ operator acting on qubit $i$. We imply tensor products with the appropriate identity factors, so $\hat{Q}$ acts on a $2^N$-dimensional Hilbert space.

\begin{figure}[tbp]
    \centering
    \includegraphics[width=0.75\columnwidth]{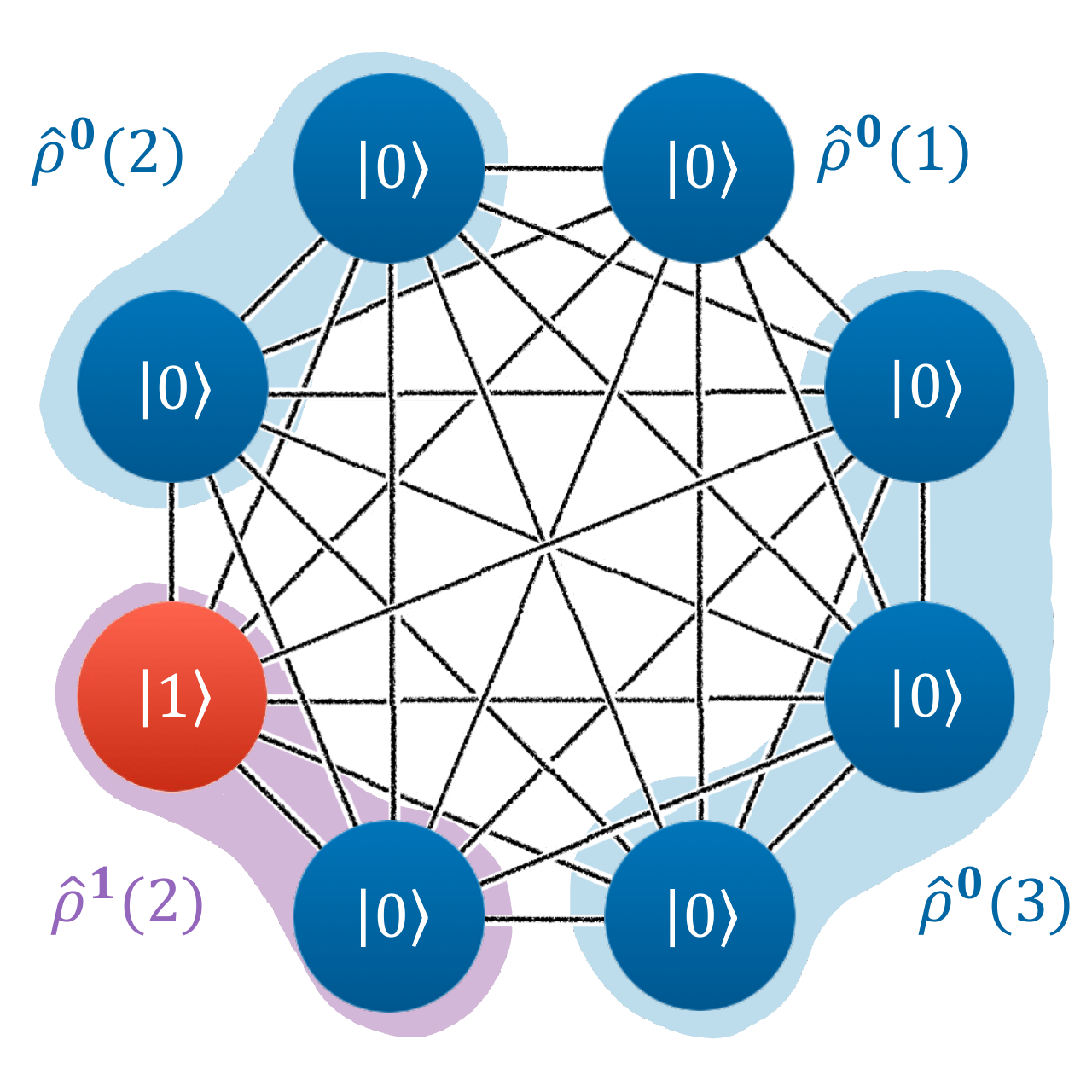}
    \caption{Dynamics along the orbit of a global pure product state with a single excitation localized on one qubit ($\ket{1}$, red), with all other qubits unexcited ($\ket{0}$, blue). Shading and superscripts distinguish two dynamical classes with distinct open dynamics. $\hat{\rho}^{\mathbf{1}}(t;K)$ is the reduced state of a $K$-qubit subsystem containing the excited qubit, while $\hat{\rho}^{\mathbf{0}}(t;K)$ excludes that qubit.}
    \label{fig:InitialState}
\end{figure}

We consider dynamics in the orbit of the generating state
\begin{equation}\label{eq:gen_state}
\hat{\rho}_{\rm gen} = \ket{\psi_{\rm gen}}\bra{\psi_{\rm gen}}, 
\qquad \ket{\psi_{\rm gen}} = \ket{100\cdots0},
\end{equation}
where $\ket{\psi_{\rm gen}}$ is a product state with a single excitation localized on one qubit. Since the network evolves unitarily, the generating state and the Hamiltonian fully determine the orbit, and the description requires no initial time or initial state. Under the all-to-all coupling, excitation disperses from the single excited qubit into the remaining $N-1$ qubits and subsequently returns. Whether a subsystem contains the excited qubit distinguishes two dynamical classes, labeled by $\mathbf{i}\in\{\mathbf{0},\mathbf{1}\}$ and illustrated in Fig.~\ref{fig:InitialState}. Class~$\mathbf{1}$ subsystems contain the excited qubit, described by reduced state $\hat{\rho}^{\mathbf{1}}(t;K)$, and class~$\mathbf{0}$ subsystems exclude that qubit, described by $\hat{\rho}^{\mathbf{0}}(t;K)$.

The Hamiltonian and generating state fix a particular tensor factorization of the global $2^N$-dimensional Hilbert space into $N$ qubits. We consider only subsystems aligned with this factorization. We set $\hat{\rho}(t=0)=\hat{\rho}_{\rm gen}$ purely as a labeling convention for points on the orbit, without privileging any initial time.

Conservation of $q$ forbids transitions between excitation sectors, so the global unitary is block diagonal in the excitation basis with blocks of dimension $\binom{N}{q}\times\binom{N}{q}$:
\begin{equation}
    \hat{U}(t)=
    \begin{pmatrix}
        u_0 &&&&&&&&& \\
        & u_s & u_d & \cdots & u_d &&&&& \\
        & u_d & u_s & \cdots & u_d &&&&& \\
        & \vdots & \vdots & \ddots & \vdots &&&&& \\
        & u_d & u_d & \cdots & u_s &&&&& \\
        &&&&& u_2 & u_{2d} & \cdots & & \\
        &&&&& u_{2d} & u_2 & \cdots & & \\
        &&&&& \vdots & \vdots & \ddots & & \\
        &&&&&&&& \ddots & \\
        &&&&&&&&& u_N
    \end{pmatrix}.
\end{equation}
Here $u_0(t)$ is the phase acquired by the ground state $\ket{00\cdots0}$ under evolution, and $u_2,\ldots,u_N$ are the diagonal amplitudes in the higher excitation sectors the orbit never occupies. Under the restriction of Sec.~\ref{sec:singleChargeSector} each of $u_2,\ldots,u_N$, like $u_0$, is a phase of modulus one. Neither $u_0$ nor the higher-sector amplitudes affect the excitation flow studied below, so only the $q=1$ block matters for those results. Within this block $u_s(t)$ denotes the amplitude for the excitation to remain on the same qubit and $u_d(t)$ denotes the transition amplitude between different qubits. Unitarity of the $q=1$ block of $\hat{U}(t)$ imposes
\begin{equation}\label{eq:unitarity_constraints}
\begin{gathered}
    |u_s|^2 + (N-1)|u_d|^2 = 1, \\
    2\Re\!\left[u_s^* u_d\right] + (N-2)|u_d|^2 = 0.
\end{gathered}
\end{equation}
Hence $u_s(t)$ and $u_d(t)$ are not independent, so the single-excitation dynamics can be expressed in terms of $u_d(t)$ alone. All subsystem propagators follow from $u_s(t)$ and $u_d(t)$, and $\left|u_d(t)\right|^2$ alone controls the excitation flow and the positivity and complete positivity of every propagator.

For the full XXZ Hamiltonian,
\begin{equation}
\hat{H}=\frac{1}{4}\sum_{i\neq j}\Big[J\big(\hat{\sigma}^x_i\hat{\sigma}^x_j+\hat{\sigma}^y_i\hat{\sigma}^y_j\big) + J_z \hat{\sigma}^z_i \hat{\sigma}^z_j\Big]+\sum_i h\hat{\sigma}^z_i,
\label{eq:XXZ_hamiltonian}
\end{equation}
the $J_z$ and $h$ terms contribute only an overall phase in each excitation sector. The $q=1$ dynamics are therefore effectively an XX model. For the homogeneous all-to-all coupling of Eq.~\eqref{eq:XXZ_hamiltonian}, we diagonalize this sector in Appendix~\ref{app:single} and obtain compact expressions for the amplitudes
\begin{equation}
u_s=\frac{1+(N-1)e^{iNJt}}{N}, \qquad
u_d=\frac{1-e^{iNJt}}{N}.
\label{eq:us_ud}
\end{equation}
Under the global unitary, $\ket{\psi_{\rm gen}}$ evolves to
\begin{equation}\label{eq:global_state}
\ket{\psi(t)} = u_s\ket{10\cdots0}+u_d\ket{01\cdots0}
+\cdots+u_d\ket{00\cdots1}.
\end{equation}

\section{Open Dynamics}\label{sec:open-dynamics}

The propagator $\Phi(t_1,t_2)$ describes the open dynamics of a $K$-qubit subsystem between times $t_1$ and $t_2$. Setting $t_1=0$ gives the dynamical map $\Phi(0,t)$. The map is CP~\cite{Choi1975,Jamiolkowski1972} because the global state $\hat{\rho}_{\rm gen}$ factorizes across all $N$ qubits at $t=0$. The joint state of any $K$-qubit subsystem and the rest of the network is also a product state. For $t_1>0$ the global state has evolved away from $\hat{\rho}_{\rm gen}$, building up correlations between each subsystem and the rest of the network that can render $\Phi(t_1,t_2)$ nonpositive and non-CP~\cite{Pechukas1994,Shaji2005,Rodriguez2010}. 

When $\Phi(t_1,t_2)$ is not CP, it can be written as the difference of two CP maps in the operator-sum representation~\cite{Sudarshan1961,dePillis1967,Jordan2004}:
\begin{equation}
    \Phi\!\left(t_1,t_2\right)\left[\hat{\rho}\right]
    =\sum_{\varphi_i>0}\varphi_i\,\hat{A}_i\,\hat{\rho}\,\hat{A}_i^\dagger
    -\sum_{\varphi_j<0}\left(-\varphi_j\right)\hat{B}_j\,\hat{\rho}\,\hat{B}_j^\dagger
\end{equation}
where $\{\varphi_i(t_1,t_2)\}$ are real coefficients grouped by sign for a particular choice of $(t_1,t_2)$, and $\{\hat{A}_i(t_1,t_2),\hat{B}_j(t_1,t_2)\}$ are excitation-basis operators that map $\hat{\rho}(t_1)$ to $\hat{\rho}(t_2)$. Because the grouping by sign changes with $(t_1,t_2)$ in our example, this form does not provide a single unified representation valid across all time windows. 

Absorbing the coefficients into the operators avoids this. We define
\begin{equation}
    \hat{\phi}_i = \sqrt{\varphi_i}\hat{A}_i, \qquad
    \hat{\phi}_j = \sqrt{\varphi_j}\hat{B}_j,
\end{equation}
so that, combining both sums, the propagator admits one operator-sum representation valid for any $(t_1,t_2)$, with the minimum number of terms:
\begin{equation}
\label{eq:MSR}
    \Phi\!\left(t_1,t_2\right)\left[\hat{\rho}\right]
    =\hat{\phi}\,\hat{\rho}\,\hat{\phi}^{\dagger}+\sum_i\hat{\phi}_i\,\hat{\rho}\,\hat{\phi}_i^T\,.
\end{equation}
The contribution $\hat{\phi}\,\hat{\rho}\,\hat{\phi}^\dagger$ is always CP. The operators $\hat{\phi}_i$ are off-block-diagonal and mediate excitation flow between excitation sectors, with their contributions $\hat{\phi}_i\,\hat{\rho}\,\hat{\phi}_i^T$ violating complete positivity when $\varphi_i(t_1,t_2)<0$. When every $\varphi_i$ vanishes only the $\dagger$ term remains and trace preservation makes $\hat{\phi}$ unitary, so the $\varphi_i$ measure the departure of the propagator from unitary dynamics. Although Eq.~\eqref{eq:MSR} resembles a Kraus decomposition, the two terms differ. The $\dagger$ term always contributes positively since $\hat{\phi}\hat{\phi}^\dagger \succeq 0$, while the $T$ term uses the matrix transpose without complex conjugation, so $(\sqrt{\varphi_i})^2$ can be negative when $\varphi_i < 0$, allowing the $T$ term to subtract rather than add. This representation minimizes the number of nonzero operators $\hat{\phi}_i(t_1,t_2)$. Other decompositions of the same propagator are possible.

The generating state $\hat{\rho}(0) = \ket{100\cdots0}\bra{100\cdots0}$ is a product state in the excitation basis, so the partial trace involves no cross terms between different basis states. The CPTP dynamical map $\Phi(0,t)$, constructed in Appendix~\ref{app:maps}, is calculated from
\begin{equation}
    \hat{\rho}_S(t)
    = \Phi(0,t)\left[\hat{\rho}_S(0)\right]
    = \operatorname{Tr}_E\!\left[\hat{U}(t)\,\hat{\rho}(0)\,
    \hat{U}^\dagger(t)\right],
\label{eq:reduced_map}
\end{equation}
where $S$ is the subsystem and $\operatorname{Tr}_E$ denotes the partial trace over $E$, the $N-K$ qubits forming the rest of the network. When $\Phi(0,t)$ is invertible, the propagator factors as
\begin{equation}
    \Phi(t_1,t_2)=\Phi(0,t_2)\,\Phi^{-1}(0,t_1).
\label{eq:compose}
\end{equation}

The two dynamical classes introduced in Sec.~\ref{sec:nqubit} give rise to structurally distinct propagators. We derive $\Phi^{\mathbf{1}}$ for subsystems containing the excited qubit and $\Phi^{\mathbf{0}}$ for those that exclude that qubit.

\subsection{Open Dynamics Restricted to the Single-Excitation Sector}\label{sec:singleChargeSector}

A subsystem density matrix carries any excitation number $q\leq K$, while every global state in the orbit of the generating state carries global excitation number one. We derive the open dynamics of Eq.~\eqref{eq:compose}, with the Hamiltonian of Eq.~\eqref{eq:XXZ_hamiltonian}, restricted to the single-excitation sector. The restriction suits a local observer who knows that the total excitation number is conserved and equal to one, and who therefore cares only about states within that sector. The restricted set of states is still larger than the orbit itself.

We require every global sector of excitation number two or more to evolve by a phase of modulus one, so that maps and propagators allow no transition between those sectors and the single-excitation sector. The convention makes the general form of the propagator tractable at arbitrary subsystem size, and it ties excitation flow to positivity and complete positivity on the restricted domain.

The detailed derivations of the CPTP maps and the propagators, restricted to the single-excitation sector, are given in Appendix~\ref{app:maps} and Appendix~\ref{app:prop}.

\subsection{Subsystem Containing the Excited Qubit}\label{sec:env-zero}

We decompose the open-subsystem propagator $\Phi^{\mathbf{1}}\!\left(t_1,t_2;K\right)$ for a $K$-qubit subsystem containing the excited qubit into a block-diagonal part, diagonal in excitation number, and an off-block-diagonal part that mediates excitation flow. The block-diagonal operator $\hat{\phi}^{\mathbf{1}}$ has blocks of size $\binom{K}{q}\times\binom{K}{q}$ for each excitation sector $q$, while the off-block-diagonal operator $\hat{\phi}_\tau^{\mathbf{1}}$ couples the $q=1$ and $q=0$ sectors. We write
\begin{equation}
\label{eq:Phi1}
\Phi^{\mathbf{1}}\!\left(t_1,t_2;K\right)\left[\hat{\rho}\right]
=\hat{\phi}^{\mathbf{1}}\hat{\rho}\,\hat{\phi}^{\mathbf{1}\dagger}
+\hat{\phi}^{\mathbf{1}}_\tau\,\hat{\rho}\,\hat{\phi}^{\mathbf{1}T}_{\tau},
\end{equation}
where the matrix elements of $\hat{\phi}^{\mathbf{1}}$ and $\hat{\phi}^{\mathbf{1}}_\tau$ are nonunitary amplitudes connecting states either within a given excitation sector or between sectors.

The block-diagonal operator is
\begin{equation}
\hat{\phi}^{\mathbf{1}}=\begin{pmatrix}
    \varphi_0^{\mathbf{1}} & & & & & & & \\
    & \varphi_s^{\mathbf{1}} & \varphi_d^{\mathbf{1}} & \cdots & \varphi_d^{\mathbf{1}} & & & \\
    & \varphi_d^{\mathbf{1}} & \varphi_s^{\mathbf{1}} & \cdots & \varphi_d^{\mathbf{1}} & & & \\
    & \vdots & \vdots & \ddots & \vdots & & & \\
    & \varphi_d^{\mathbf{1}} & \varphi_d^{\mathbf{1}} & \cdots & \varphi_s^{\mathbf{1}} & & & \\
    & & & & & \ddots & & \\
    & & & & & & \varphi_q^{\mathbf{1}} & \\
    & & & & & & & \ddots
\end{pmatrix},
\end{equation}
where the $q=1$ sector contains two amplitudes, $\varphi_s^{\mathbf{1}}$ and $\varphi_d^{\mathbf{1}}$. Here $\varphi_s^{\mathbf{1}}$ is the nonunitary amplitude for the excitation to remain on the same qubit, and $\varphi_d^{\mathbf{1}}$ is the amplitude for transitions between different qubits within the subsystem. These play the same role for the open-subsystem propagator as $u_s$ and $u_d$ play for the global unitary. Together they govern the internal redistribution of excitation, from the excited qubit to the remaining $K-1$ unexcited qubits inside the subsystem. The symbols $\varphi_q^{\mathbf{1}}$ with $q\neq1$ are the assigned phases of Sec.~\ref{sec:singleChargeSector}, of modulus one and otherwise unconstrained. In terms of the global unitary,
\begin{equation}
\begin{gathered}
\varphi_s^{\mathbf{1}}=\frac{u_d\!\left(t_1\right)u_d\!\left(t_2\right)-u_s\!\left(t_1\right)u_s\!\left(t_2\right)}{\big[u_d\!\left(t_1\right)-u_s\!\left(t_1\right)\big]\big[(K-1)u_d\!\left(t_1\right)+u_s\!\left(t_1\right)\big]}\\
\hspace{2em}+\frac{(K-2)u_d\!\left(t_1\right)\big[u_d\!\left(t_2\right)-u_s\!\left(t_2\right)\big]}{\big[u_d\!\left(t_1\right)-u_s\!\left(t_1\right)\big]\big[(K-1)u_d\!\left(t_1\right)+u_s\!\left(t_1\right)\big]},\\
\varphi_d^{\mathbf{1}}=\frac{u_d\!\left(t_1\right)u_s\!\left(t_2\right)-u_s\!\left(t_1\right)u_d\!\left(t_2\right)}{\big[u_d\!\left(t_1\right)-u_s\!\left(t_1\right)\big]\big[(K-1)u_d\!\left(t_1\right)+u_s\!\left(t_1\right)\big]},\\
\left|\varphi_q^{\mathbf{1}}\right|=1, \qquad q\in\{0,2,3,\ldots,K\}.
\end{gathered}
\label{eq:phi_sdq1}
\end{equation}

The off-block-diagonal operator is
\begin{equation}\label{eq:phitau_op1}
\hat{\phi}_\tau^{\mathbf{1}}=\left(
\begin{array}{ccccccccc}
\cline{1-1}
\multicolumn{1}{|c|}{q=0} & \sqrt{\varphi_\tau^{\mathbf{1}}} & \cdots & \sqrt{\varphi_\tau^{\mathbf{1}}} & & & & & \\
\cline{1-1}\cline{2-4}
 & \multicolumn{3}{|c|}{} & & & & & \\
 & \multicolumn{3}{|c|}{\rule{0pt}{2.5ex}} & & & & & \\
 & \multicolumn{3}{|c|}{q=1} & & & & & \\
 & \multicolumn{3}{|c|}{\rule[-2.5ex]{0pt}{0pt}} & & & & & \\
 & \multicolumn{3}{|c|}{} & & & & & \\
\cline{2-4}
 & & & & & & & & \\
 & & & & & & & & \\
 & & & & & & & & \\
 & & & & & & & & \phantom{0000}
\end{array}
\right),
\end{equation}
where $\sqrt{\varphi_\tau^{\mathbf{1}}}=\bra{q=0}\hat{\phi}_\tau^{\mathbf{1}}\ket{q=1}$ is either real or purely imaginary. All other entries are zero. The boxes indicate the $q=0$ and $q=1$ excitation sectors to orient the reader. The $K$ nonzero entries $\sqrt{\varphi_\tau^{\mathbf{1}}}$ occupy the single $q=0$ row across all $K$ columns of the $q=1$ block, reflecting that $\hat{\phi}_\tau^{\mathbf{1}}$ mediates transitions between the $q=0$ sector and each of the $K$ states in the $q=1$ block. This structure places the dynamics in the form of Eq.~\eqref{eq:MSR}. In terms of the global unitary,
\begin{equation}\label{eq:phi_tau1}
\varphi_\tau^{\mathbf{1}}
= \frac{\left|u_d\!\left(t_2\right)\right|^2-\left|u_d\!\left(t_1\right)\right|^2}
       {\frac{1}{N-K}-K\!\left|u_d\!\left(t_1\right)\right|^2}.
\end{equation}
The numerator fixes the sign of $\varphi_\tau^{\mathbf{1}}$, since the denominator is always positive. Here $\varphi_\tau^{\mathbf{1}}>0$ indicates net dispersal away from the excited qubit, while $\varphi_\tau^{\mathbf{1}}<0$ indicates net backflow. This sign controls positivity and complete positivity. At $K=N$ the subsystem is the whole network, no excitation leaves it, and $\varphi_\tau^{\mathbf{1}}(t_1,t_2;N)=0$ reduces the propagator to the unitary evolution $\Phi^{\mathbf{1}}(t_1,t_2;N)\!\left[\hat{\rho}\right] = \hat{U}\hat{\rho}\,\hat{U}^\dagger$.

Unitarity of the full global network and excitation conservation [Eq.~\eqref{eq:unitarity_constraints}] constrain the propagator elements:
\begin{equation}\label{eq:trace_pres_1}
\begin{gathered}
    |\varphi_s^{\mathbf{1}}|^2 + (K-1)\,|\varphi_d^{\mathbf{1}}|^2 + \varphi_\tau^{\mathbf{1}} = 1,\\
    2\Re\!\left[\varphi_s^{\mathbf{1}^*} \varphi_d^{\mathbf{1}}\right] + (K-2)|\varphi_d^{\mathbf{1}}|^2 + \varphi_\tau^{\mathbf{1}} = 0.
\end{gathered}
\end{equation}
These relations enforce trace preservation and constrain all three of $\varphi_s^{\mathbf{1}}$, $\varphi_d^{\mathbf{1}}$, and $\varphi_\tau^{\mathbf{1}}$. Equivalently,
\begin{equation}
\hat{\phi}^{\mathbf{1}\dagger}\hat{\phi}^{\mathbf{1}}
+\hat{\phi}^{\mathbf{1}T}_{\tau}\hat{\phi}^{\mathbf{1}}_\tau=\hat{\mathbbm{1}}.
\end{equation}

\subsection{Subsystem Excluding the Excited Qubit}\label{sec:env-one}

We apply the decomposition of Eq.~\eqref{eq:MSR} to the open-subsystem propagator $\Phi^{\mathbf{0}}\!\left(t_1,t_2;K\right)$ for a $K$-qubit subsystem that excludes the excited qubit. The propagator separates into two block-diagonal pieces, $\hat{\phi}^{\mathbf{0}}$ and $\hat{\phi}^{\mathbf{0}}_0$, together with an off-block-diagonal piece $\hat{\phi}_\tau^{\mathbf{0}}$ that mediates excitation flow:
\begin{equation}
\label{eq:Phi0}
\Phi^{\mathbf{0}}\left(t_1,t_2;K\right)\left[\hat{\rho}\right]
=\hat{\phi}^{\mathbf{0}}\hat{\rho}\,\hat{\phi}^{\mathbf{0}\dagger}
+\hat{\phi}^{\mathbf{0}}_0\,\hat{\rho}\,\hat{\phi}^{\mathbf{0}T}_{0}
+\hat{\phi}^{\mathbf{0}}_\tau\,\hat{\rho}\,\hat{\phi}^{\mathbf{0}T}_{\tau}.
\end{equation}

The first block-diagonal operator is
\begin{equation}
\hat{\phi}^{\mathbf{0}}=\begin{pmatrix}
    \varphi_s^{\mathbf{0}} & & & & & & & \\
    & \varphi_2^{\mathbf{0}} & & & & & & \\
    & & \varphi_2^{\mathbf{0}} & & & & & \\
    & & & \ddots & & & & \\
    & & & & \varphi_2^{\mathbf{0}} & & & \\
    & & & & & \ddots & & \\
    & & & & & & \varphi_{q+1}^{\mathbf{0}} & \\
    & & & & & & & \ddots
\end{pmatrix},
\end{equation}
\begin{equation}\label{eq:phi_sq0}
\begin{gathered}
\varphi_s^{\mathbf{0}}=\frac{u_s\!\left(t_2\right)}{u_s\!\left(t_1\right)},\\
\left|\varphi_{q+1}^{\mathbf{0}}\right|=1,
\qquad q\in\{1,2,\ldots,K\},
\end{gathered}
\end{equation}
where $\varphi_s^{\mathbf{0}}$ is a nonunitary amplitude for the $K$-qubit ground state to remain in the ground state. Notably, there is no analog of $\varphi_d^{\mathbf{1}}$ here. The $K$ qubits of the class~$\mathbf{0}$ subsystem are unexcited and homogeneously coupled, so no qubit is distinguishable from another and there is no internal mixing of excitation within the subsystem. The symbols $\varphi_{q+1}^{\mathbf{0}}$ are the assigned phases of Sec.~\ref{sec:singleChargeSector}, carrying the shifted subscript because a local sector of excitation number $q$ sits in the global sector of excitation number $q+1$.

The second block-diagonal operator has a single nonzero entry:
\begin{equation}
\hat{\phi}_0^{\mathbf{0}}=\begin{pmatrix}
    \sqrt{\varphi_0^{\mathbf{0}}-\left|\varphi_s^{\mathbf{0}}\right|^2} & \phantom{\sqrt{\varphi_\tau^{\mathbf{0}}}} & & & & & & & & \\
    & & & & & & & & & \\
    & & & & & & & & & \\
    & & & & & & & & & \\
    & & & & & & & & & \\
    & & & & & & & & & \\
    & & & & & & & & & \\
    & & & & & & & & & \\
    & & & & & & & & & \\
    & & & & & & & & & 
\end{pmatrix},
\end{equation}
where $\sqrt{\varphi_0^{\mathbf{0}}-\left|\varphi_s^{\mathbf{0}}\right|^2}=\bra{q=0}\hat{\phi}_0^{\mathbf{0}}\ket{q=0}$ is either real or purely imaginary. In terms of the global unitary,
\begin{equation}\label{eq:phi_00}
\varphi_0^{\mathbf{0}}=\frac{1-K\left|u_d\!\left(t_2\right)\right|^2}{1-K\left|u_d\!\left(t_1\right)\right|^2}.
\end{equation}

The off-block-diagonal operator is
\begin{equation}\label{eq:phitau_op0}
\hat{\phi}_\tau^{\mathbf{0}}=\left(
\begin{array}{ccccccccc}
\cline{1-1}
\multicolumn{1}{|c|}{q=0} & & & & & & & & \\
\cline{1-1}\cline{2-6}
\rule{0pt}{3.5ex}\sqrt{\varphi_\tau^{\mathbf{0}}} & \multicolumn{5}{|c|}{} & & & \\
\vdots & \multicolumn{5}{|c|}{q=1} & & & \\
\rule[-1.5ex]{0pt}{4.5ex}\sqrt{\varphi_\tau^{\mathbf{0}}} & \multicolumn{5}{|c|}{\phantom{00000000000}} & & & \\
\cline{2-6}
 & & & & & & & & \\
 & & & & & & & & \\
 & & & & & & & & \\
 & & & & & & & & \phantom{0000000}
\end{array}
\right),
\end{equation}
where $\sqrt{\varphi_\tau^{\mathbf{0}}}=\bra{q=1}\hat{\phi}_\tau^{\mathbf{0}}\ket{q=0}$ is either real or purely imaginary. All other entries are zero. The boxes indicate the $q=0$ and $q=1$ excitation sectors to orient the reader. The $K$ nonzero entries $\sqrt{\varphi_\tau^{\mathbf{0}}}$ occupy the single $q=0$ column across all $K$ rows of the $q=1$ block, reflecting that $\hat{\phi}_\tau^{\mathbf{0}}$ mediates transitions between the $q=0$ sector and each of the $K$ states in the $q=1$ block. In terms of the global unitary,
\begin{equation}\label{eq:phi_tau0}
\varphi_\tau^{\mathbf{0}}=\frac{\left|u_d\!\left(t_2\right)\right|^2-\left|u_d\!\left(t_1\right)\right|^2}{1-K\left|u_d\!\left(t_1\right)\right|^2}.
\end{equation}
The numerator controls the sign of $\varphi_\tau^{\mathbf{0}}$, since the denominator is always positive. Unlike the sign convention for $\varphi_\tau^{\mathbf{1}}$, $\varphi_\tau^{\mathbf{0}}>0$ denotes net flow \emph{into} the class~$\mathbf{0}$ subsystem, while $\varphi_\tau^{\mathbf{0}}<0$ denotes net backflow out of that subsystem. In both cases the sign of $\varphi_\tau^{\mathbf{i}}$ encodes the direction of net excitation flow in a unified way, with $\varphi_\tau^{\mathbf{i}}>0$ corresponding to dispersal away from the excited qubit and $\varphi_\tau^{\mathbf{i}}<0$ to backflow.

Unitarity of the full global network and excitation conservation [Eq.~\eqref{eq:unitarity_constraints}] constrain the propagator elements:
\begin{equation}\label{eq:trace_pres_0}
    \varphi_0^{\mathbf{0}} + K\,\varphi_\tau^{\mathbf{0}} = 1.
\end{equation}
This constraint enforces trace preservation. Equivalently,
\begin{equation}
\hat{\phi}^{\mathbf{0}\dagger}\hat{\phi}^{\mathbf{0}}
+\hat{\phi}^{\mathbf{0}T}_0\hat{\phi}^{\mathbf{0}}_0+\hat{\phi}^{\mathbf{0}T}_\tau\hat{\phi}^{\mathbf{0}}_\tau=\hat{\mathbbm{1}}.
\end{equation}

\subsection{Excitation Flow}\label{sec:exciteflow}

\begin{figure}[tbp]
    \centering
    \includegraphics[width=\columnwidth]{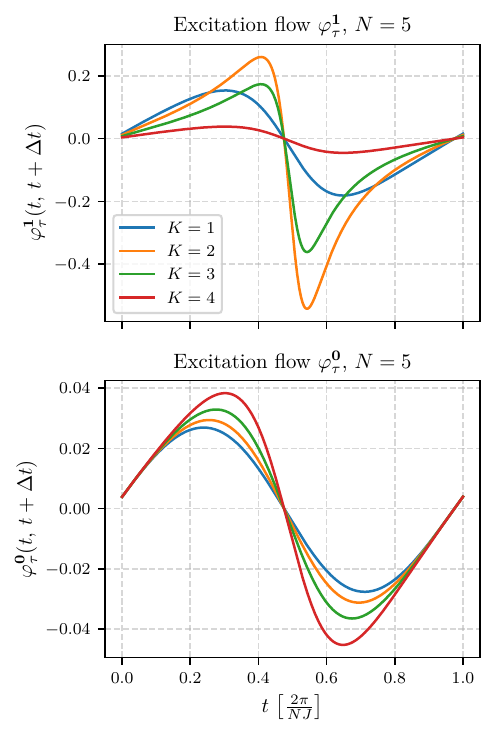}
    \caption{Excitation-flow parameter $\varphi_\tau^{\mathbf{i}}(t,t+\Delta t;K)$ [Eqs.~\eqref{eq:phi_tau1},~\eqref{eq:phi_tau0}] for $\Delta t=0.05$, $K\in\{1,2,3,4\}$, and $N=5$, plotted against $t$ in units of $2\pi/(NJ)$. Positive values indicate net dispersal away from the excited qubit, where the dynamics are positive and CP. Negative values indicate backflow, where the dynamics are neither.}
    \label{fig:tau_combined}
\end{figure}

The phases of Sec.~\ref{sec:singleChargeSector} exclude transitions into the sectors of excitation number $q\geq2$, so excitation flows only between the $q=0$ and $q=1$ sectors and one excitation-flow parameter in each class, $\varphi_\tau^{\mathbf{i}}$, measures the flow. The sign of $\varphi_\tau^{\mathbf{i}}(t_1,t_2;K)$ gives the direction of excitation flow between the subsystems. Figure~\ref{fig:tau_combined} plots $\varphi_\tau^{\mathbf{i}}(t,t+\Delta t;K)$ against $t$ for $\Delta t=0.05$, $K\in\{1,2,3,4\}$, and $N=5$, with $t$ in units of $2\pi/(NJ)$.

The excitation-flow parameter satisfies $\varphi_{\tau}^{\mathbf{i}}(t,t+\Delta t;K)>0$ for $t<0.5-\Delta t/2=0.475$, where the net excitation flow over $[t,t+\Delta t]$ is still outward. For $t>0.475$ the sign reverses and excitation backflows toward the excited qubit, driving the state back toward the generating state.

The single excited qubit shares excitation equally among $N-1$ recipients, each receiving a fraction of order $|u_d|^2 \sim 4/N^2$. For fixed $K$, $\varphi_\tau^{\mathbf{1}}$ scales as $4/N$ because the class~$\mathbf{1}$ subsystem contains the excited qubit and loses to all $N-K$ qubits outside it, while $\varphi_\tau^{\mathbf{0}}$ scales as $4/N^2$ since each qubit receives only its individual share. For fixed $N$, $\varphi_\tau^{\mathbf{1}}$ decreases with $K$ since fewer qubits remain outside to receive excitation, while $\varphi_\tau^{\mathbf{0}}$ increases with $K$ since more qubits are available inside to receive excitation.

Combining the expressions for $\varphi_\tau^{\mathbf{1}}(t_1,t_2;K)$ and $\varphi_\tau^{\mathbf{0}}(t_1,t_2;K)$ yields the conservation relation
\begin{equation}\label{eq:conserve_tau}
    \left(\left|u_d\left(t_2\right)\right|^2-\left|u_d\left(t_1\right)\right|^2\right)
    \left(\frac{1}{\varphi_\tau^{\mathbf{0}}}-\frac{1}{\varphi_\tau^{\mathbf{1}}}\right)
    = 1 - \frac{1}{N-K}.
\end{equation}
This relation encodes excitation balance between class~$\mathbf{1}$ and class~$\mathbf{0}$ subsystems and holds for any two $K$-qubit subsystems of the same network over the same time window $[t_1,t_2]$, and the pair need not span the full network. The equal-size restriction simplifies the denominators of Eqs.~\eqref{eq:phi_tau1} and~\eqref{eq:phi_tau0}, and the generalization to unequal sizes involves additional terms.

\section{Reduced States}\label{sec:reduced-states}

The reduced states of both dynamical classes have rank two for all $K$, since the single-excitation structure of the global state [Eq.~\eqref{eq:global_state}] leaves each reduced state as a mixture of one $q=0$ and one $q=1$ vector. The reduced states take the form 
\begin{equation}\label{eq:reduced_states}
\begin{split}
\hat{\rho}^{\mathbf{1}}(t;K) &= \left(1-p^{\mathbf{1}}\right)
\ket{00\cdots0}\bra{00\cdots0} + p^{\mathbf{1}}
\ket{\psi^{\mathbf{1}}}\bra{\psi^{\mathbf{1}}},\\
\hat{\rho}^{\mathbf{0}}(t;K) &= 
p^{\mathbf{0}}\ket{00\cdots0}\bra{00\cdots0} + \left(1-p^{\mathbf{0}}\right)
\ket{\psi^{\mathbf{0}}}\bra{\psi^{\mathbf{0}}},
\end{split}
\end{equation}
where $p^{\mathbf{i}}(t;K)$ and $\ket{\psi^{\mathbf{i}}(t;K)}$ are time-dependent. We suppress the arguments below for brevity. For dynamical class~$\mathbf{1}$,
\begin{gather}
\ket{\psi^{\mathbf{1}}} = \frac{u_s\ket{10\cdots0}+u_d\ket{01\cdots0}
+\cdots+u_d\ket{00\cdots1}}{\sqrt{p^{\mathbf{1}}}},\nonumber\\
\begin{aligned}
\label{eq:p1}
p^{\mathbf{1}}(t;K) &= |u_s(t)|^2+(K-1)|u_d(t)|^2 \\
               &= 1 - \frac{4(N-K)}{N^2}
               \sin^2\!\left(\frac{NJt}{2}\right),
\end{aligned}
\end{gather}
and for dynamical class~$\mathbf{0}$,
\begin{equation}
\begin{gathered}
\ket{\psi^{\mathbf{0}}} = \frac{\ket{10\cdots0}+\ket{01\cdots0}
+\cdots+\ket{00\cdots1}}{\sqrt{K}},\\
\begin{aligned}
\label{eq:p0}
p^{\mathbf{0}}(t;K) &= 1-K|u_d(t)|^2 \\
               &= 1-\frac{4K}{N^2}
               \sin^2\!\left(\frac{NJt}{2}\right).
\end{aligned}
\end{gathered}
\end{equation}
The two weights satisfy $p^{\mathbf{1}}(t;K) = p^{\mathbf{0}}(t;N-K)$, reflecting conservation of total excitation. Since the global state is pure, all statistical mixing is a consequence of partitioning into subsystems. For a partition of the network into a single class~$\mathbf{1}$ subsystem of size $K_1$ and class~$\mathbf{0}$ subsystems of sizes $K_i$ with $i>1$,
\begin{equation}
\begin{gathered}
1-p^{\mathbf{1}}(t;K_1) = \sum_{i>1}\left[1-p^{\mathbf{0}}(t;K_i)\right],\\ 
\sum_i K_i = N.
\end{gathered}
\end{equation}
The probability of finding the excitation outside the class~$\mathbf{1}$ subsystem equals the sum of the probabilities of finding the excitation inside each class~$\mathbf{0}$ subsystem, a direct consequence of single-excitation conservation.

The limits $K=N$ and $K=1$ bracket the range of dynamical behavior. When $K=N$, $p^{\mathbf{1}}(t;N)=1$ for all $t$ and $\hat{\rho}^{\mathbf{1}}(t;N) = \ket{\psi(t)}\bra{\psi(t)}$ is the pure global state [Eq.~\eqref{eq:global_state}], with all dynamics residing in the evolving state vector. At $K=1$, the reduced state is diagonal for all $t$ and the dynamics reduce to excitation flow between $\ket{0}$ and $\ket{1}$, with the eigenvectors $\{\ket{0},\ket{1}\}$ fixed throughout. The coherence dynamics encoded in $\varphi_d^{\mathbf{1}}$ in Sec.~\ref{sec:env-zero} are absent at $K=1$ and become accessible only for $K\geq 2$.

The structure of $\hat{\rho}^{\mathbf{1}}$ directly reflects the propagator decomposition of Sec.~\ref{sec:open-dynamics}. The excitation probability $p^{\mathbf{1}}$ is set by excitation flow across the subsystem boundary, controlled by $\varphi_\tau^{\mathbf{1}}$, while the internal state $\ket{\psi^{\mathbf{1}}}$ is set by internal redistribution among the $K$ qubits, controlled by $\varphi_s^{\mathbf{1}}$ and $\varphi_d^{\mathbf{1}}$. For class~$\mathbf{0}$, the ground-state probability $p^{\mathbf{0}}$ is likewise set by excitation flow across the boundary, controlled by $\varphi_\tau^{\mathbf{0}}$, while $\ket{\psi^{\mathbf{0}}}$ is time-independent since all $K$ qubits are equivalent and there is no internal redistribution.

\section{\texorpdfstring{Positivity, Fixed Points,\\and Trace Distance}{Positivity, Fixed Points, and Trace Distance}}\label{sec:pcp}

Since any density operator admits a spectral decomposition $\hat{\rho} = \sum_a p_a \ket{\psi_a}\bra{\psi_a}$ with $p_a \geq 0$, linearity of $\Phi^{\mathbf{i}}$ reduces positivity to the requirement that $\Phi^{\mathbf{i}}$ map every pure state $\ket{\psi_a}\bra{\psi_a}$ to a positive operator. Testing $\Phi^{\mathbf{i}}(t_1,t_2;K)$ on $K$-qubit pure states therefore yields the positivity criterion
\begin{equation}\label{eq:positivity}
    \Phi^{\mathbf{i}}(t_1,t_2;K)\ \text{is positive}
    \quad\Longleftrightarrow\quad
    \varphi_\tau^{\mathbf{i}}(t_1,t_2;K) \ge 0.
\end{equation}
Propagators of any subsystem size and either dynamical class are positive when excitation disperses away from the excited qubit and nonpositive when excitation backflows.

Complete positivity is a stronger requirement in general. It holds if and only if the Choi matrix~\cite{Choi1975,Jamiolkowski1972},
\begin{equation}
    \hat{C}_\Phi = \sum_{\mu,\nu} \Phi^{\mathbf{i}}\Big[ \ket{\mu}\bra{\nu} \Big] \otimes \ket{\mu}\bra{\nu},
\end{equation}
is positive semidefinite. Appendix~\ref{app:choi} computes its spectrum and finds one eigenvalue that can be negative for class~$\mathbf{1}$ and two for class~$\mathbf{0}$, each negative when and only when $\varphi_\tau^{\mathbf{i}}<0$, so
\begin{equation}\label{eq:CP}
    \Phi^{\mathbf{i}}(t_1,t_2;K)\ \text{is CP}
    \quad\Longleftrightarrow
\quad
    \varphi_\tau^{\mathbf{i}}(t_1,t_2;K) \geq 0\,.
\end{equation}
Positivity and complete positivity therefore coincide for every subsystem size and both dynamical classes, and no propagator is positive without also being CP. The $q=0$ sector holds the $K$-qubit ground state alone, so the single $q=0$ row of $\hat{\phi}_\tau^{\mathbf{1}}$ [Eq.~\eqref{eq:phitau_op1}] and the single $q=0$ column of $\hat{\phi}_\tau^{\mathbf{0}}$ [Eq.~\eqref{eq:phitau_op0}] leave both operators at rank one for every $K$, and that rank is why Eqs.~\eqref{eq:positivity} and~\eqref{eq:CP} share one inequality. Two conditions keep the rank at one, the restriction of Sec.~\ref{sec:singleChargeSector} on the sectors of excitation number $q\geq2$ and the single excitation of the generating state.

Removing either condition opens the flow to sectors above $q=1$. Lifting the restriction while keeping the generating state of Eq.~\eqref{eq:gen_state} separates positivity and complete positivity for class~$\mathbf{0}$ at $K=1$, verified at $N=4$ for homogeneous all-to-all and nearest-neighbor ring couplings. Replacing the generating state by a state carrying two excitations, with the sectors $q\geq3$ evolving by a phase, separates positivity and complete positivity for class~$\mathbf{1}$ at $K=1$, verified at $N\leq6$. With both conditions in place, inhomogeneous couplings $J_{ij}$ and inhomogeneous local fields $h_i$ change the nonzero entries of $\hat{\phi}_\tau^{\mathbf{i}}$ without adding a row or column. Positivity and complete positivity still coincide, verified at $N\leq7$ and $K\leq5$.

Studies of CP dynamics, especially in the master equation setting, often assume zero-discord initial states, a sufficient but not necessary condition~\cite{Pechukas1994,Shaji2005,Rodriguez2010,PerezGarcia2006,Shabani2009,Modi2012b,Brodutch2013,Colla2022,Chruscinski2022}. Maps and propagators are two-time quantities, however, so both the initial correlations and the dynamics fix their positivity and complete positivity~\cite{Jagadish2023}. We instead use the contractivity properties of positive maps to connect the sign of $\varphi_\tau^{\mathbf{i}}$ to the expansion or contraction of the trace norm. 

A propagator is nonpositive if and only if there exists a Hermitian operator whose trace norm strictly expands under the propagator~\cite{Kossakowski1972RMP,Kossakowski1972Bull,Wissmann:2015}. Each propagator in this model has a fixed point $\hat{\rho}^{\mathbf{i}}_{\rm fix}$ shared by its class, so we test this criterion on the operators $\hat{\rho}^{\mathbf{i}}(t)-\hat{\rho}^{\mathbf{i}}_{\rm fix}$. At $K=1$ each class has one fixed point, and for $K>1$ each has a fixed manifold. For class~$\mathbf{1}$ the fixed points are the ground state $\ket{00\cdots0}$, where $p^{\mathbf{1}}_{\rm fix} = 0$ [Eq.~\eqref{eq:reduced_states}], together with the local $q=1$ states whose amplitudes satisfy $\sum_i c_i = 0$. The block-diagonal operator $\hat{\phi}^{\mathbf{1}}$ gives these states a phase, while $\hat{\phi}_\tau^{\mathbf{1}}$ contributes nothing, so the propagator returns them unchanged. For class~$\mathbf{0}$ the fixed points are the entire local $q=1$ sector. The block-diagonal operator $\hat{\phi}^{\mathbf{0}}$ gives these states a phase, while $\hat{\phi}_0^{\mathbf{0}}$ and $\hat{\phi}_\tau^{\mathbf{0}}$ contribute nothing, so the propagator returns them unchanged.

Unlike the class~$\mathbf{1}$ manifold, this degeneracy follows from the restriction of Sec.~\ref{sec:singleChargeSector}, since it leaves $\varphi_2^{\mathbf{0}}$ a phase and excludes nonunitary terms analogous to $\varphi_d^{\mathbf{1}}$. Lifting the restriction for that sector replaces $\varphi_2^{\mathbf{0}}$ by a block that mixes states within the manifold, so the manifold shrinks and varies with the time window rather than being shared by every class~$\mathbf{0}$ propagator.

The reduced state $\hat{\rho}^{\mathbf{i}}$ has rank two with eigenvalues $p^{\mathbf{i}}$ and $1-p^{\mathbf{i}}$, and $\hat{\rho}^{\mathbf{i}}_{\rm fix}$ projects onto the eigenvector with eigenvalue $1-p^{\mathbf{i}}$. The weight $p^{\mathbf{i}}(t;K)$ is therefore the trace distance from the fixed point, $\frac{1}{2}\|\hat{\rho}^{\mathbf{i}} - \hat{\rho}^{\mathbf{i}}_{\rm fix}\|_1 = p^{\mathbf{i}}$, for any $K$. Substituting into Eqs.~\eqref{eq:phi_tau1} and~\eqref{eq:phi_tau0} expresses $\varphi_\tau^{\mathbf{i}}$ in terms of the trace distances,
\begin{equation}
\begin{aligned}
\varphi_\tau^{\mathbf{1}} &= 
\frac{p^{\mathbf{1}}(t_1;K) - p^{\mathbf{1}}(t_2;K)}
{1 - K\left[1-p^{\mathbf{1}}(t_1;K)\right]}, \\
\varphi_\tau^{\mathbf{0}} &= 
\frac{p^{\mathbf{0}}(t_1;K) - p^{\mathbf{0}}(t_2;K)}
{K\, p^{\mathbf{0}}(t_1;K)}.
\end{aligned}
\end{equation}
Both denominators are positive, so whether the trace distance to the fixed point increases or decreases over $[t_1,t_2]$ alone fixes the sign of $\varphi_\tau^{\mathbf{i}}$. Equations~\eqref{eq:positivity} and~\eqref{eq:CP} then yield the trace distance criterion
\begin{equation}\label{eq:pop_criterion}
\begin{gathered}
\Phi^{\mathbf{i}}(t_1,t_2;K)\ \text{is positive and CP}\\
\iff p^{\mathbf{i}}(t_2;K) \leq p^{\mathbf{i}}(t_1;K).
\end{gathered}
\end{equation}
A propagator is therefore positive and CP if and only if the propagator contracts the state toward the fixed point, and nonpositive and non-CP if and only if the propagator moves the state away. The nonmonotonicity of these trace distances and the failure of the dynamical map to be positive divisible signal the non-Markovian nature of these dynamics~\cite{Breuer2009,Rivas2010,Wissmann:2015}.

\section{Ensemble of Single-Qubit Propagators}\label{sec:geometry}

Specializing to single-qubit subsystems with $K=1$ allows us to characterize the ensemble of propagators $\{\Phi^{\mathbf{i}}(t_1,t_2;1)\}$ geometrically and to identify several features that may persist in less symmetric settings. The Bloch vector $\vec{b}$ in the unit Bloch ball represents a single-qubit density operator~\cite{Bengtsson2017}, with the ground state $\ket{0}$ at $b_z=+1$ and the excited state $\ket{1}$ at $b_z=-1$.

We express the single-qubit propagators of Eqs.~\eqref{eq:Phi1} and~\eqref{eq:Phi0} in the Pauli basis as the matrices $\hat{\Lambda}^{\mathbf{1}}(t_1,t_2)$ and $\hat{\Lambda}^{\mathbf{0}}(t_1,t_2)$ acting on Bloch vectors, with the hat marking a matrix representation of the propagator $\Phi^{\mathbf{i}}$:
\begin{gather}
    \hat{\Lambda}^{\mathbf{1}}(t_1,t_2)\,\vec{b} =
    \begin{pmatrix}
        1 &   &   &   \\[4pt]
          & \lvert \varphi^{\mathbf{1}}_s \rvert c_{\theta^{\mathbf{1}}} & -\lvert \varphi^{\mathbf{1}}_s \rvert s_{\theta^{\mathbf{1}}} &   \\[4pt]
          & \lvert \varphi^{\mathbf{1}}_s \rvert s_{\theta^{\mathbf{1}}} & \lvert \varphi^{\mathbf{1}}_s \rvert c_{\theta^{\mathbf{1}}} &   \\[4pt]
        \varphi^{\mathbf{1}}_\tau &   &   & \lvert \varphi^{\mathbf{1}}_s \rvert^2
    \end{pmatrix}
    \begin{pmatrix}\label{eq:Lambda1}
        1 \\[4pt] b_x \\[4pt] b_y \\[4pt] b_z
    \end{pmatrix}, \nonumber\\
    \varphi^{\mathbf{1}}_s = \frac{u_s(t_2)}{u_s(t_1)}, 
    \qquad 
    \varphi^{\mathbf{1}}_\tau = 1 - \lvert \varphi^{\mathbf{1}}_s \rvert^2,
\end{gather}
where $c_{\theta^{\mathbf{1}}}\equiv\cos{\theta^{\mathbf{1}}}$, $s_{\theta^{\mathbf{1}}}\equiv\sin{\theta^{\mathbf{1}}}$, and ${\theta^{\mathbf{1}}}=\arg\!\left[\varphi^{\mathbf{1}*}_0\varphi^{\mathbf{1}}_s\right]$ is the rotation angle in the $XY$ plane. The angle enters no result below.
\begin{gather}
    \hat{\Lambda}^{\mathbf{0}}(t_1,t_2)\,\vec{b} =
    \begin{pmatrix}
        1 &   &   &   \\[4pt]
          & \lvert \varphi^{\mathbf{0}}_s \rvert c_{\theta^{\mathbf{0}}} & -\lvert \varphi^{\mathbf{0}}_s \rvert s_{\theta^{\mathbf{0}}} &   \\[4pt]
          & \lvert \varphi^{\mathbf{0}}_s \rvert s_{\theta^{\mathbf{0}}} & \lvert \varphi^{\mathbf{0}}_s \rvert c_{\theta^{\mathbf{0}}} &   \\[4pt]
        -\varphi^{\mathbf{0}}_\tau &   &   & \varphi^{\mathbf{0}}_0
    \end{pmatrix}
    \begin{pmatrix}\label{eq:Lambda0}
        1 \\[4pt] b_x \\[4pt] b_y \\[4pt] b_z
    \end{pmatrix}, \nonumber\\
    \varphi^{\mathbf{0}}_0 = 
    \frac{1 - \lvert u_d(t_2) \rvert^2}{1 - \lvert u_d(t_1) \rvert^2}, 
    \qquad 
    \varphi^{\mathbf{0}}_\tau = 1 - \varphi^{\mathbf{0}}_0, \nonumber\\
    \varphi^{\mathbf{0}}_s=\frac{u_s\!\left(t_2\right)}{u_s\!\left(t_1\right)},
\end{gather}
where ${\theta^{\mathbf{0}}}=\arg{\left[\varphi^{\mathbf{0}*}_s\varphi^{\mathbf{0}}_2\right]}$, with $c_{\theta^{\mathbf{0}}}$ and $s_{\theta^{\mathbf{0}}}$ defined analogously. The block-diagonal entries rescale the Bloch components, while the excitation-flow parameter $\varphi_\tau^{\mathbf{i}}$ shifts the $z$-component. The fixed point of $\hat{\Lambda}^{\mathbf{1}}(t_1,t_2)$ is the north pole $b_z=+1$, the ground state $\ket{0}$, and the fixed point of $\hat{\Lambda}^{\mathbf{0}}(t_1,t_2)$ is the south pole $b_z=-1$, the excited state $\ket{1}$. 

The time-dependent distributions of individual propagator parameters and the time-independent distribution of fixed points offer useful benchmarks for comparison to other ensembles. The symmetry of the generating state and the homogeneous coupling ensure that $\varphi_s^{\mathbf{i}}$, $\varphi_0^{\mathbf{0}}$, and $\varphi_\tau^{\mathbf{i}}$ evolve with a common frequency $NJ/(2\pi)$. In less symmetric settings the parameters evolve with several frequencies~\cite{Prudhoe2024}. Within a given dynamical class, the fixed point of every $\hat{\Lambda}^{\mathbf{i}}(t_1,t_2)$ sits at the same pole of the Bloch ball regardless of $t_1$ and $t_2$, in marked contrast to systems that thermalize or otherwise relax subsystem states toward a steady state. A numerical circuit model gives time-dependent distributions of both propagator parameters and fixed points~\cite{Akhouri2025}.

We next characterize the propagator ensemble by comparing the domain of positivity of each single-qubit propagator to the reduced states $\hat{\rho}^{\mathbf{i}}(t;1)$. Although the dynamical map $\hat{\Lambda}^{\mathbf{i}}(0,t)$ always contracts the Bloch ball toward the fixed point, the propagators $\hat{\Lambda}^{\mathbf{i}}(t_1,t_2)$ can either contract or expand the ball, depending on the sign of $\varphi_\tau^{\mathbf{i}}(t_1,t_2)$. We call a propagator with $\varphi_\tau^{\mathbf{i}}(t_1,t_2)>0$ \emph{contracting} and a propagator with $\varphi_\tau^{\mathbf{i}}(t_1,t_2)<0$ \emph{expanding}~\cite{PerezGarcia2006,Chruscinski2022,Jagadish2023}. A contracting propagator decreases the ball volume and moves every input state closer to the fixed point. An expanding propagator increases the volume and moves every input state farther away. Contraction corresponds to CP dynamics and expansion to non-CP dynamics, consistent with Eq.~\eqref{eq:CP}.

\begin{figure}[tbp]
    \centering
    \includegraphics[width=\columnwidth]{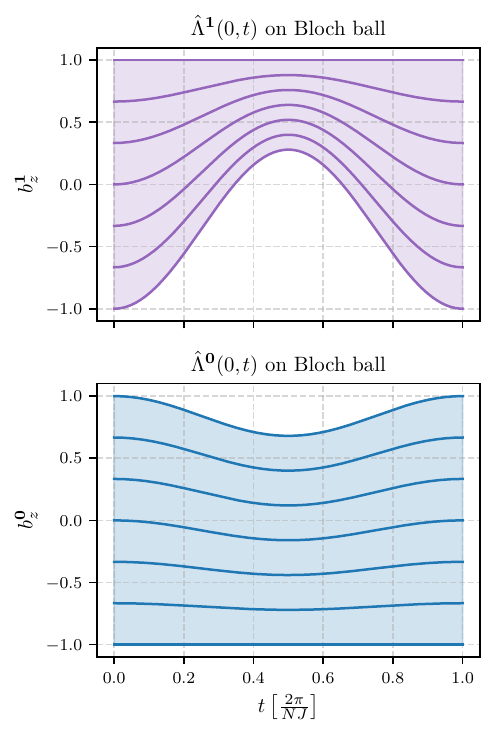}
    \caption{Action of the single-qubit dynamical map $\hat{\Lambda}^{\mathbf{i}}(0,t)$ 
    on the $z$-component of the Bloch ball for $N=5$, with 
    $b_z(0)\in\{-1,-2/3,-1/3,0,1/3,2/3,1\}$. Propagators 
    $\hat{\Lambda}^{\mathbf{i}}(t_1,t_2)$ contract the Bloch ball toward the fixed point 
    when $\varphi_\tau^{\mathbf{i}}>0$ and expand the ball away from it 
    when $\varphi_\tau^{\mathbf{i}}<0$.}
    \label{fig:azLambda_combined}
\end{figure}

Figure~\ref{fig:azLambda_combined} illustrates the action of the dynamical maps $\hat{\Lambda}^{\mathbf{i}}(0,t)$ on the $z$-component of the Bloch ball, where $\vec{b}^{\mathbf{i}}(t)=\hat{\Lambda}^{\mathbf{i}}(0,t)\vec{b}(0)$ with $b_z(0) \in [-1,1]$.
Explicitly,
\begin{equation}
\begin{gathered}
    b_z^{\mathbf{1}}(t) = \lvert \varphi^{\mathbf{1}}_s(0,t) \rvert^2\, b_z(0) 
        + \varphi^{\mathbf{1}}_\tau(0,t), \\
    b_z^{\mathbf{0}}(t) = \varphi^{\mathbf{0}}_0(0,t)\, b_z(0) 
        - \varphi^{\mathbf{0}}_\tau(0,t),
\end{gathered}
\end{equation}
where
\begin{align}
\begin{split}
    \varphi^{\mathbf{1}}_\tau(0,t) 
        &=1-\lvert u_s(t) \rvert^2 
        = \frac{4(N-1)}{N^2}
           \sin^2\!\left(\frac{NJt}{2}\right),\\
    \varphi^{\mathbf{0}}_\tau(0,t) 
        &= \lvert u_d(t) \rvert^2 
        = \frac{4}{N^2}
           \sin^2\!\left(\frac{NJt}{2}\right).
\end{split}
\end{align}
Both fixed points are visible as the convergence points of the trajectories during the first half-period. $\hat{\Lambda}^{\mathbf{1}}(0,t)$ drives all states toward $b_z=+1$, while $\hat{\Lambda}^{\mathbf{0}}(0,t)$ drives all states toward $b_z=-1$. Both contraction and expansion weaken as $N$ increases because larger networks dilute the excitation flow into each subsystem, reducing the magnitude of $\varphi_\tau^{\mathbf{i}}$.

Figure~\ref{fig:azLambda_combined} also shows the trajectory of the reduced state $\hat{\rho}^{\mathbf{i}}(t;1)$. For class~$\mathbf{1}$, the trajectory is the curve starting at the south pole $b_z^{\mathbf{1}}(0)=-1$, the excited state $\ket{1}$, and moving upward toward the fixed point as excitation disperses. For class~$\mathbf{0}$, the trajectory is the curve starting at the north pole $b_z^{\mathbf{0}}(0)=+1$, the ground state $\ket{0}$, and moving downward as excitation flows into the subsystem.

A nonpositive propagator maps only a restricted set of input states to valid states. For $\hat{\Lambda}^{\mathbf{i}}(t_1,t_2)$ this set, the domain of positivity, is the subset of Bloch vectors $\vec{d}^{\,\mathbf{i}}$ whose image remains inside the Bloch ball,
\begin{equation}
    \mathcal{P}^{\mathbf{i}}(t_1,t_2)\equiv \left\{\,\vec{d}^{\,\mathbf{i}}\;\middle|\;\bigl\|\hat{\Lambda}^{\mathbf{i}}(t_1,t_2)\,\vec{d}^{\,\mathbf{i}}\bigr\| \le 1\,\right\},
\end{equation}
where the norm is that of the three-vector part of the image.

\begin{figure}[tbp]
    \centering
    \includegraphics[width=\columnwidth]{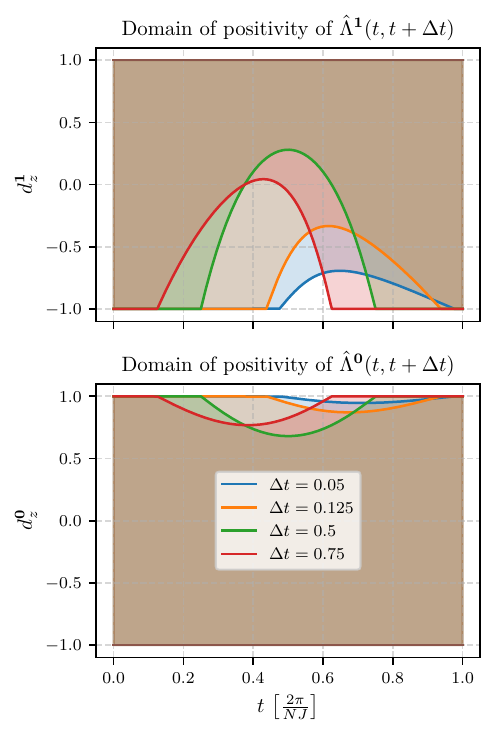}
    \caption{Shaded regions show the $z$-component of the domain of positivity of $\hat{\Lambda}^{\mathbf{i}}(t,t+\Delta t)$ with $N=5$. Contracting propagators preserve positivity across the full Bloch interval $d^{\mathbf{i}}_z \in [-1,1]$, whereas expanding propagators preserve positivity only on a restricted subset. Overlapping shades appear brown.}
    \label{fig:pz_combined}
\end{figure}

Figure~\ref{fig:pz_combined} shows the $z$-component of this domain for propagators over time windows of fixed duration $\Delta t$, $\hat{\Lambda}^{\mathbf{i}}(t,t+\Delta t)$ with $\Delta t\in\{0.05,0.125,0.5,0.75\}$ in units of $2\pi/(NJ)$, evaluated with $N=5$. The short-time propagators $\hat{\Lambda}^{\mathbf{i}}(t,t+0.05)$ are contracting for $t\in[0,0.475]$, where $\varphi_\tau^{\mathbf{i}}(t,t+0.05)>0$ in Fig.~\ref{fig:tau_combined}, and therefore positive, preserving the full Bloch interval $d^{\mathbf{i}}_z\in[-1,1]$. For $t\in[0.475,0.975]$ the same short-time propagators are expanding and become nonpositive, preserving positivity only for a restricted set of input states. That set lies above the blue curve for $\hat{\Lambda}^{\mathbf{1}}$ and below the blue curve for $\hat{\Lambda}^{\mathbf{0}}$.

As $\Delta t$ increases, the net flow over the time window turns from dispersal to backflow at an earlier $t_1$, so the positive regime shifts toward earlier $t$. Consequently $\hat{\Lambda}^{\mathbf{i}}(t,t+0.125)$ becomes expanding at earlier $t$ than $\hat{\Lambda}^{\mathbf{i}}(t,t+0.05)$, so its domain of positivity excludes input states earlier.

We compare the domain of positivity of each single-qubit propagator with the reduced states $\hat{\rho}^{\mathbf{i}}(t;1)$, shown in Fig.~\ref{fig:azLambda_combined} as the curves starting at $b_z^{\mathbf{1}}(0)=-1$ and $b_z^{\mathbf{0}}(0)=+1$. An expanding $\Phi^{\mathbf{0}}(t_1,t_2;1)$ applied to the class~$\mathbf{0}$ reduced state $\hat{\rho}^{\mathbf{0}}(t;1)$ at a time $t\neq t_1$ takes the state outside the Bloch ball. A class~$\mathbf{0}$ propagator therefore does not preserve positivity along the entire class~$\mathbf{0}$ trajectory. By contrast, $\Phi^{\mathbf{0}}(t_1,t_2;1)$ preserves positivity for any $t_1$ and $t_2$ when applied to the class~$\mathbf{1}$ reduced state $\hat{\rho}^{\mathbf{1}}(t;1)$, and $\Phi^{\mathbf{1}}(t_1,t_2;1)$ preserves positivity for any $t_1$ and $t_2$ when applied to the class~$\mathbf{0}$ reduced state $\hat{\rho}^{\mathbf{0}}(t;1)$.

A band of states lies inside the domain of positivity of every propagator $\Phi^{\mathbf{i}}(t_1,t_2;1)$, yet the reduced state never visits these states. The band has no counterpart in positive- and CP-divisible dynamics, where any propagator may be applied to any state at any time.

For $K=1$, the eigenvalues of the density operator are $(1\pm|\vec{b}|)/2$, so positivity reduces to the single condition $|\hat{\Lambda}^{\mathbf{i}}(t_1,t_2)\,\vec{b}|\leq 1$. This condition permits an explicit characterization of the domain. For $K>1$ the positivity conditions involve the full density matrix structure, including local Bloch vectors, the correlation matrix, its determinant, and mixed norms~\cite{Gamel2016}, making the domain of positivity substantially harder to characterize.

\section{\texorpdfstring{Entanglement Entropy\\and Positivity}{Entanglement Entropy and Positivity}}\label{sec:subsystem-correlations}

The direction of change of the trace distance to the fixed point controls the positivity structure of Secs.~\ref{sec:pcp} and~\ref{sec:geometry}. We calculate the quantum correlations between each $K$-qubit subsystem and the rest of the network as a function of time.

The quantum discord quantifies correlations beyond entanglement~\cite{Ollivier2001,HendersonVedral2001,Dillenschneider2008,Werlang2010,Ferraro2010,RulliSarandy2011,Liu2011,PalBose2011,Campbell2011a,Campbell2011b,Saguia2011,Tomasello2011,Werlang2011,Tomasello2012,Allegra2011,Maziero2012,LiLin2011,Dhar2012,PalBose2012a,PalBose2012b,Mazzola2010,Xu2010a,Prabhu2012b,Modi2012a},
\begin{equation}
    \mathcal{D}_{S|E} = S\!\left[\hat{\rho}_S\right] - S\!\left[\hat{\rho}\right] 
    + \min_{\{\hat{\Pi}^k_S\}} S\!\left[\hat{\rho}_{E|\{\hat{\Pi}^k_S\}}\right],
\end{equation}
where $S[\hat{\rho}]=-\mathrm{Tr}[\hat{\rho}\ln\hat{\rho}]$ is the von Neumann entropy and the minimization runs over all local projective measurements on the subsystem. For the global pure state considered here, $S[\hat{\rho}]=0$ and a measurement on the subsystem in the Schmidt basis leaves the rest of the network in a pure state, so the minimum conditional entropy also vanishes. The quantum discord therefore equals the entanglement entropy,
\begin{equation}\label{eq:entropySE}
\mathcal{D}_{S|E}=S\!\left[\hat{\rho}_S\right]=S\!\left[\hat{\rho}_E\right],
\end{equation}
and all bipartite correlations are entanglement.

Since the reduced state has rank two for all $K$, the entanglement entropy depends only on $p^{\mathbf{i}}(t;K)$. From Eqs.~\eqref{eq:reduced_states}, \eqref{eq:p1}, and~\eqref{eq:p0}, the entanglement entropies evaluate to
\begin{align}
    S\!\left[\hat{\rho}^{\mathbf{1}}\right]
        &= - (N-K)\,|u_d|^2 \ln\!\big[(N-K)\,|u_d|^2\big] \nonumber\\
        &\quad - \Big[1 - (N-K)|u_d|^2\Big]\,
                \ln\!\Big[1 - (N-K)|u_d|^2\Big], \nonumber\\
    S\!\left[\hat{\rho}^{\mathbf{0}}\right]
        &= - K\,|u_d|^2 \ln\!\big[K|u_d|^2\big] \\
        &\quad - \Big[1 - K|u_d|^2\Big]\,
                \ln\!\Big[1 - K|u_d|^2\Big]. \nonumber
\end{align}
Since $p^{\mathbf{0}}(t;K) = p^{\mathbf{1}}(t;N-K)$, a class~$\mathbf{0}$ subsystem and the rest of the network, a class~$\mathbf{1}$ subsystem, have equal entanglement entropies:
\begin{equation}
    S\!\left[\hat{\rho}^{\mathbf{0}}(K)\right]
        = S\!\left[\hat{\rho}^{\mathbf{1}}(N-K)\right],
\end{equation}
confirming the second equality in Eq.~\eqref{eq:entropySE}.

The $K$ dependence mirrors that of the excitation-flow parameters in Sec.~\ref{sec:exciteflow}, with class~$\mathbf{1}$ entropy decreasing with $K$ while class~$\mathbf{0}$ entropy increases. Both entropies also decrease with $N$, since larger networks dilute the excitation and reduce mixing in every subsystem.

\begin{figure}[tbp]
    \centering
    \includegraphics[width=\columnwidth]{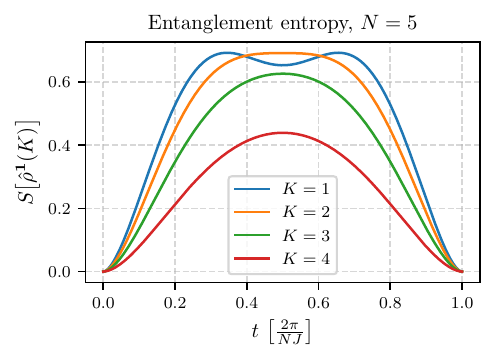}
    \caption{Entanglement entropy $S\!\left[\hat{\rho}^{\mathbf{1}}(t;K)\right]$ plotted against $t$ in units of $2\pi/(NJ)$ for $K\in\{1,2,3,4\}$ and $N=5$.}
    \label{fig:entropy_plot}
\end{figure}

Figure~\ref{fig:entropy_plot} plots $S\!\left[\hat{\rho}^{\mathbf{1}}(K)\right]$ for $K\in\{1,2,3,4\}$ and $N=5$.
At $t=0$, the global state factorizes across all qubits and quantum correlations vanish. As $|u_d|^2$ grows and the excitation disperses away from the excited qubit, bipartite correlations grow and local states become more mixed. For $K\in\{2,3,4\}$ in Fig.~\ref{fig:entropy_plot}, $p^{\mathbf{1}}(t;K)$ remains above $1/2$ for all $t$, so correlations grow monotonically until half the period and then decrease symmetrically, returning to zero after one full period. For $K=1$, $p^{\mathbf{1}}$ dips below $1/2$ at half-period, as the curve starting at $b_z^{\mathbf{1}}(0)=-1$ in Fig.~\ref{fig:azLambda_combined} shows. The reduced state passes through the maximally mixed state at $b_z=0$ and continues toward the fixed point, so the entropy dips rather than peaks at half-period. Notably, the propagator is still contracting, with $\varphi_\tau^{\mathbf{1}}>0$, as the state crosses the equator, yet the state moves toward greater purity. Contraction does not imply increasing entropy once the state has passed the maximally mixed point.

The quantum correlations and the positivity of the propagator both depend on $|u_d(t)|^2$, but in fundamentally different ways. Positivity depends on the \emph{direction of change} of $|u_d(t)|^2$ over $[t_1,t_2]$, with propagators CP while $|u_d|^2$ increases and the excitation disperses, and non-CP while $|u_d|^2$ decreases and the excitation returns. The correlations $\mathcal{D}_{S|E} = S\!\left[\hat{\rho}_S\right]$, by contrast, depend nonmonotonically on the \emph{value} of $|u_d(t)|^2$, with the entropy maximized at $p^{\mathbf{i}} = 1/2$ and decreasing on either side. For subsystems large enough that $p^{\mathbf{i}}$ remains above $1/2$ throughout, correlations grow while $|u_d|^2$ increases and shrink while $|u_d|^2$ decreases. CP propagators therefore accompany growing correlations in this regime. For smaller subsystems where $p^{\mathbf{i}}$ dips below $1/2$ during the first half-period, correlations peak before $|u_d|^2$ reaches its maximum and then begin shrinking, even while propagators are still CP. In all cases, the direction of change of the trace distance to the fixed point alone determines positivity, not the correlations at $t_1$ or their change over $[t_1,t_2]$.

\section{Inference with Open Systems}
\label{sec:complementarity}

The propagator structure derived in the preceding sections has direct operational consequences for inference. As a first example, comparing the excitation-flow parameters $\varphi_\tau^{\mathbf{1}}$ and $\varphi_\tau^{\mathbf{0}}$ from class~$\mathbf{1}$ and class~$\mathbf{0}$ single-qubit subsystems directly constrains the size and coupling of the rest of the network. An observer with access to a class~$\mathbf{1}$ and a class~$\mathbf{0}$ single-qubit subsystem can ask whether $\Phi^{\mathbf{1}}(t_1,t_2;1)$ and $\Phi^{\mathbf{0}}(t_1,t_2;1)$ are consistent with a closed two-qubit network, and if not, how many additional degrees of freedom a unitary dilation of the composite dynamics requires~\cite{Stinespring1955,Sudarshan1961}.

Any excitation-conserving two-qubit unitary gives
\begin{equation}
    \varphi_\tau^{\mathbf{1}}(t_1,t_2;1)=\varphi_\tau^{\mathbf{0}}(t_1,t_2;1),
\end{equation}
a direct consequence of excitation conservation. If $\varphi_\tau^{\mathbf{1}}(t_1,t_2;1) \neq \varphi_\tau^{\mathbf{0}}(t_1,t_2;1)$, the observer infers excitation exchange with additional degrees of freedom beyond the two observed qubits. Substituting the observed $\varphi_\tau^{\mathbf{1}}$, $\varphi_\tau^{\mathbf{0}}$, and $\left|u_d\right|^2=1-p^{\mathbf{0}}(t;1)$ into Eq.~\eqref{eq:conserve_tau} infers $N$, giving the number $N-2$ of additional qubits required for the network to be closed. Once $N$ is inferred, the oscillation period $2\pi/(NJ)$ directly determines $J$ for the homogeneous coupling considered here.

To go beyond this qualitative inference and quantify sensitivity to global parameters, we use the quantum Fisher information~\cite{Helstrom1976,Braunstein1994,Paris2009}, a bound on how well an optimal measurement on a quantum state determines a parameter. Much of the literature on quantum parameter estimation focuses on optimizing probe states~\cite{Giovannetti2004,Giovannetti2006}, encoding schemes~\cite{Degen2017}, or measurement strategies~\cite{Helstrom1976,Braunstein1994}. The setting here differs in two respects. The subsystems are natural partitions of a closed autonomous network rather than engineered probes, and the parameters of interest characterize the global network rather than the initial state of a probe. Because the subsystems share quantum correlations with each other and with the rest of the network, they are not independent probes and their Fisher information does not combine additively~\cite{Toth:2014,Pezze:2018}.

\subsection{Fisher Information for Global Parameters}
\label{sec:FIglobal}

We compute the quantum Fisher information for global parameters $\Theta$ from the reduced states of each subsystem, focusing on the coupling strength $J$ and the total number of qubits $N$. The symmetric logarithmic derivative $\hat{L}$ and the Fisher information $\mathcal{F}_{\Theta}$ satisfy
\begin{equation}
\begin{aligned}
\partial_\Theta \hat{\rho}(t;\Theta)
&= \frac{1}{2}
\left[
\hat{L}(t;\Theta)\,\hat{\rho}(t;\Theta)
+ \hat{\rho}(t;\Theta)\,\hat{L}(t;\Theta)
\right], \\
\mathcal{F}_{\Theta}(t)
&= \operatorname{Tr}
\left[
\hat{L}^2(t;\Theta)\,\hat{\rho}(t;\Theta)
\right].
\end{aligned}
\end{equation}
The eigenvalues $p_a$ and eigenvectors $\ket{\psi_a}$ of the reduced states $\hat{\rho}^{\mathbf{i}}(t;K)$ derived in Sec.~\ref{sec:reduced-states} determine $\hat{L}$, giving
\begin{equation}
\mathcal{F}_{\Theta}(t)
=
\sum_a \frac{\bigl(\partial_\Theta p_a\bigr)^2}{p_a}
+
\sum_{a\neq b}
\frac{2(p_a - p_b)^2}{p_a + p_b}
\left|
\braket{\psi_a | \partial_{\Theta} \psi_b}
\right|^2 .
\end{equation}
The first sum is the Fisher information of the eigenvalue distribution $\{p_a\}$, the classical contribution $\mathcal{F}_\Theta^c$, and the second, arising from the parameter dependence of the eigenvectors, is the quantum contribution $\mathcal{F}_\Theta^q$.

Evaluating $\mathcal{F}_J$ using the reduced states of Sec.~\ref{sec:reduced-states} gives
\begin{equation}
\begin{aligned}
\mathcal{F}_J^c\!\left[\hat{\rho}^{\mathbf{1}}(K)\right]
&=
4 t^2
\frac{(N-K)\,
\cos^2\!\left(\frac{NJt}{2}\right)}
{1 - 4\frac{N-K}{N^2}
\sin^2\!\left(\frac{NJt}{2}\right)}, \\
\mathcal{F}_J^q\!\left[\hat{\rho}^{\mathbf{1}}(K)\right]
&=
4 t^2
\frac{K-1}
{1 - 4\frac{N-K}{N^2}
\sin^2\!\left(\frac{NJt}{2}\right)}, \\
\mathcal{F}_J^c\!\left[\hat{\rho}^{\mathbf{0}}(K)\right]
&=
4 t^2
\frac{K\,
\cos^2\!\left(\frac{NJt}{2}\right)}
{1 - 4\frac{K}{N^2}
\sin^2\!\left(\frac{NJt}{2}\right)}, \\
\mathcal{F}_J^q\!\left[\hat{\rho}^{\mathbf{0}}(K)\right]
&= 0 .
\end{aligned}
\end{equation}

States evolved under coupling $J$ and $J+\delta J$ accumulate a relative phase $N\delta J\,t$ that grows linearly with time. Since the Fisher information scales as the square of the derivative of this phase with respect to $J$, every contribution to $\mathcal{F}_J$ carries an overall $t^2$ secular growth. The elapsed phase difference thus acts as an internal time coordinate for the global network, with longer time windows yielding greater sensitivity to $J$.

For dynamical class~$\mathbf{1}$, the limits $K=N$ and $K=1$ illustrate the classical--quantum decomposition. At $K=N$ the subsystem is the full network and $\hat{\rho}^{\mathbf{1}}(N) = \ket{\psi(t)}\bra{\psi(t)}$ is pure, evolving unitarily. Its eigenvalues do not depend on $J$, and $J$ appears only in the eigenvector $\ket{\psi(t)}$, through $u_d$, so $\mathcal{F}_J^c = 0$ and the Fisher information is purely quantum. The result $\mathcal{F}_J[\hat{\rho}^{\mathbf{1}}(N)] = 4t^2(N-1)$ combines the $t^2$ secular growth of the internal time coordinate with the $N-1$ orthogonal states the excitation can transfer into, one per unexcited qubit, each carrying the same amplitude $u_d$. Unitarity leaves $u_s$ with no independent information once $u_d$ is known. At $K=1$ the $q=1$ sector is one-dimensional, so the reduced state is diagonal with eigenvectors $\{\ket{0},\ket{1}\}$ independent of $J$. All distinguishability resides in $p^{\mathbf{1}}(t;1) = |u_s(t)|^2$, so $\mathcal{F}_J^q = 0$ and the Fisher information is purely classical.

For dynamical class~$\mathbf{0}$, the $q=1$ eigenvector $\ket{\psi^{\mathbf{0}}}$ is an equal superposition of all $q=1$ states, independent of $J$ (Sec.~\ref{sec:reduced-states}), so $\mathcal{F}_J[\hat{\rho}^{\mathbf{0}}(K)]$ is purely classical. Analogous to the weights $p^{\mathbf{0}}(t;K) = p^{\mathbf{1}}(t;N-K)$,
\begin{equation}
\mathcal{F}_J^c\!\left[\hat{\rho}^{\mathbf{0}}(K)\right]=
\mathcal{F}_J^c\!\left[\hat{\rho}^{\mathbf{1}}(N-K)\right].
\end{equation}

\begin{figure}[tbp]
    \centering
    \includegraphics[width=\columnwidth]{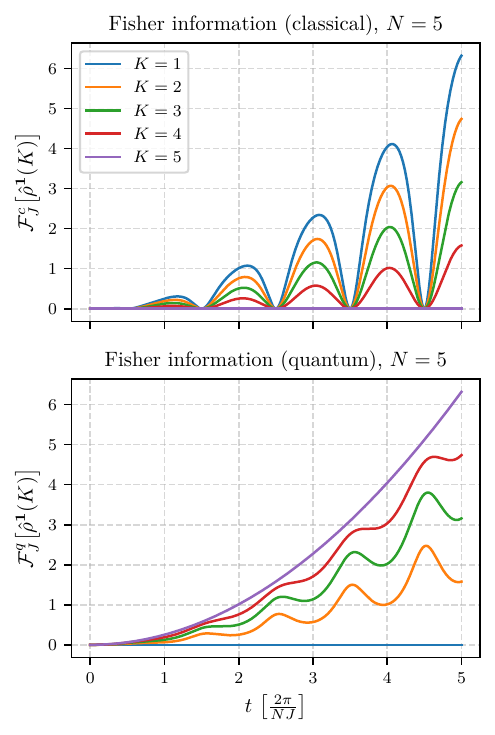}
    \caption{Classical and quantum contributions to the Fisher information 
    $\mathcal{F}_J\bigl[\hat{\rho}^{\mathbf{1}}(t;K)\bigr]$ for dynamical 
    class~$\mathbf{1}$, plotted against $t$ in units of $2\pi/(NJ)$ for 
    $K\in\{1,2,3,4,5\}$ and $N=5$.}
    \label{fig:Fisher_combined}
\end{figure}

Figure~\ref{fig:Fisher_combined} plots the classical and quantum contributions to $\mathcal{F}_J\bigl[\hat{\rho}^{\mathbf{1}}(K)\bigr]$ for $K\in\{1,2,3,4,5\}$ and $N=5$. The classical contribution peaks at integer multiples of the global period, when the excitation is most localized, and dips at half-integer multiples, when it is most spread out. The quantum contribution is complementary, peaking at half-integer multiples, when the eigenvectors are most misaligned with the excitation basis, and dipping at integer multiples, when they are perfectly aligned. As Sec.~\ref{sec:reduced-states} shows, excitation flow across the subsystem boundary sets the eigenvalues $p^{\mathbf{1}}$ and $1-p^{\mathbf{1}}$ and so drives $\mathcal{F}_\Theta^c$, while internal redistribution sets the eigenvector $\ket{\psi^{\mathbf{1}}}$ and so drives $\mathcal{F}_\Theta^q$. For all $K<N$, $\mathcal{F}_\Theta^c$ oscillates more strongly in time than $\mathcal{F}_\Theta^q$, so excitation flow across the boundary dominates the oscillation. Both share the $t^2$ secular growth.

We treat $N$ as a continuous parameter when differentiating, so that $\mathcal{F}_N$ measures sensitivity to a small change in network size. Unlike $\mathcal{F}_J$, the expressions for $\mathcal{F}_N$ contain cross terms scaling as $t$ rather than $t^2$, so the secular growth does not factor cleanly from the oscillatory modulation. Writing $\mathcal{F}_N^{c\mathbf{i}}$ and $\mathcal{F}_N^{q\mathbf{i}}$ for the classical and quantum contributions to $\mathcal{F}_N\bigl[\hat{\rho}^{\mathbf{i}}(t;K)\bigr]$,
\begin{align}
\mathcal{F}_N^{c\mathbf{1}} &= \frac{4\left[(N-2K)\sin\!\left(\frac{\alpha}{2}\right)
    -(N-K)NJt\cos\!\left(\frac{\alpha}{2}\right)\right]^2}
    {N^2(N-K)\left[N^2-4(N-K)\sin^2\!\left(\frac{\alpha}{2}\right)\right]}, \nonumber\\
\mathcal{F}_N^{q\mathbf{1}} &= \frac{4(K-1)\left[(NJt)^2
    -2NJt\sin(\alpha)+4\sin^2\!\left(\frac{\alpha}{2}\right)\right]}
    {N^2\left[N^2-4(N-K)\sin^2\!\left(\frac{\alpha}{2}\right)\right]}, \nonumber\\
\mathcal{F}_N^{c\mathbf{0}} &= \frac{4K\left[Jt\cos\!\left(\frac{\alpha}{2}\right)
    -\frac{2}{N}\sin\!\left(\frac{\alpha}{2}\right)\right]^2}
    {N^2-4K\sin^2\!\left(\frac{\alpha}{2}\right)}, \nonumber\\
\mathcal{F}_N^{q\mathbf{0}} &= 0, \quad \alpha = NJt.
\end{align}

Because the subsystems are correlated with the rest of the network, the Fisher information is nonadditive:
\begin{equation}
\mathcal{F}_\Theta\!\left[\hat{\rho}^{\mathbf{1}}(K_a)\right]
+\mathcal{F}_\Theta\!\left[\hat{\rho}^{\mathbf{0}}(K_b)\right]
\neq
\mathcal{F}_\Theta\!\left[\hat{\rho}^{\mathbf{1}}(K_a+K_b)\right].
\end{equation}
Correlations generically modify the total information accessible from a given partition in ways that cannot be determined from the individual subsystem contributions alone.

\begin{figure*}[tbp]
    \centering
    \includegraphics[width=\textwidth]{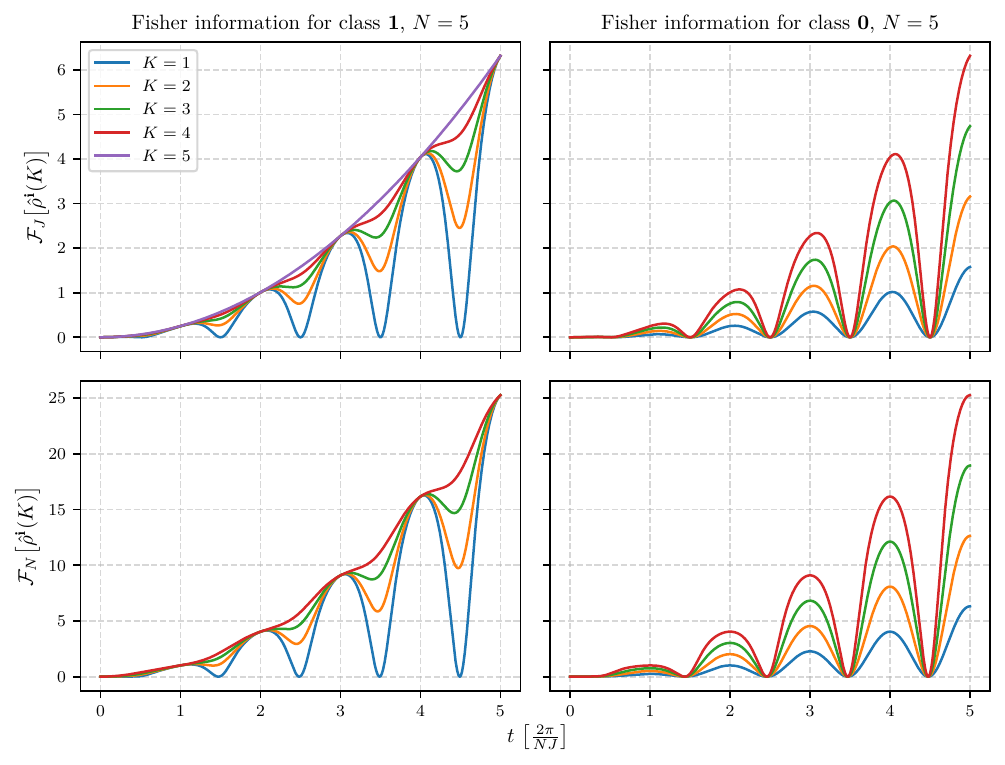}
    \caption{Fisher information $\mathcal{F}_J$ (top) and $\mathcal{F}_N$ (bottom) 
    for both dynamical classes, plotted against $t$ in units of $2\pi/(NJ)$ for $K\in\{1,2,3,4,5\}$ and $N=5$. Class~$\mathbf{0}$ is shown for $K<N$ only. $\mathcal{F}_N^{\mathbf{1}}$ diverges at $K=N$.}
    \label{fig:Fisher_total_N}
\end{figure*}

Figure~\ref{fig:Fisher_total_N} shows the total Fisher information $\mathcal{F}_\Theta = \mathcal{F}_\Theta^c + \mathcal{F}_\Theta^q$ for both dynamical classes, with class~$\mathbf{0}$ only for $K<N$ since a class~$\mathbf{0}$ subsystem excludes the excited qubit. The total grows secularly as $t^2$ with local dips at half-integer multiples of the period. At $K=1$, $\mathcal{F}_\Theta^q = 0$ and the dips reach zero, while for $K>1$ the quantum contribution keeps them above zero. $\mathcal{F}_N^{\mathbf{1}}$ diverges at $K=N$. There the observer has access to every qubit and already knows $N$, so there is nothing to infer.

\subsection{\texorpdfstring{State and Process Contributions to\\Fisher Information}{State and Process Contributions to Fisher Information}}
\label{sec:complementarity-sub}

The classical--quantum decomposition of Sec.~\ref{sec:FIglobal} identifies whether the information about global parameters resides in the eigenvalue distribution or in the eigenvectors of $\hat{\rho}$. An observer generally begins observing at some $t_1>0$, not at $t=0$, so the subsystem state at $t_1$ is not the reduced state of $\hat{\rho}_{\rm gen}$. We therefore decompose the Fisher information into a portion carried by the state at $t_1$ and a portion generated by the propagator over $[t_1,t_2]$.

A parameter governing global evolution will generically appear both in the state at $t_1$ and in the propagator over $[t_1,t_2]$. The parameter dependence of the state at $t_2$ then decomposes into process and state contributions as
\begin{equation}
\begin{split}
    \partial_{\Theta}\hat{\rho}(t_2;\Theta)
    &= \partial_{\Theta}\!\left(\Phi(t_1,t_2;\Theta)\Bigl[\hat{\rho}(t_1;\Theta)\Bigr]\right)\\
    &= \underbrace{\Bigl(\partial_{\Theta}\Phi(t_1,t_2;\Theta)\Bigr)}_{\text{process}}\Bigl[\hat{\rho}(t_1;\Theta)\Bigr] \\
    &\quad+ \Phi(t_1,t_2;\Theta)\!\underbrace{\Bigl[\partial_{\Theta}\hat{\rho}(t_1;\Theta)\Bigr]}_{\text{state}}.
\end{split}
\end{equation}

For $K=1$ the reduced state is diagonal for all $t$ and the weight $p^{\mathbf{i}}(t)$ characterizes the state completely, so only the classical contribution survives. The propagator equations reduce at $K=1$ to the multiplicative form
\begin{equation}
p^{\mathbf{i}}(t_2;\Theta) = \Bigl(1-\varphi_\tau^{\mathbf{i}}(t_1,t_2;\Theta)\Bigr)p^{\mathbf{i}}(t_1;\Theta),
\end{equation}
giving, after differentiation with respect to $\Theta$,
\begin{equation}
\partial_{\Theta} p^{\mathbf{i}}(t_2)
= \underbrace{-\partial_{\Theta}\varphi_\tau^{\mathbf{i}}}_{\text{process}}
\, p^{\mathbf{i}}(t_1) 
+ \left(1-\varphi_\tau^{\mathbf{i}}\right)
\underbrace{\partial_{\Theta} p^{\mathbf{i}}(t_1)}_{\text{state}}.
\end{equation}
The Fisher information decomposes as
\begin{align}
\mathcal{F}_\Theta^{\mathbf{i}} &= \frac{1}{p^{\mathbf{i}}(t_2)(1-p^{\mathbf{i}}(t_2))}
\Bigl[
\underbrace{\left(\partial_\Theta\varphi_\tau^{\mathbf{i}}\right)^2 \left(p^{\mathbf{i}}(t_1)\right)^2}_{\text{process}} \nonumber\\
&\quad- \underbrace{2\,\partial_\Theta\varphi_\tau^{\mathbf{i}}\, p^{\mathbf{i}}(t_1) \left(1-\varphi_\tau^{\mathbf{i}}\right) \partial_\Theta p^{\mathbf{i}}(t_1)}_{\text{cross}} \nonumber\\
&\quad+\underbrace{\left(1-\varphi_\tau^{\mathbf{i}}\right)^2\left(\partial_\Theta p^{\mathbf{i}}(t_1)\right)^2}_{\text{state}}
\Bigr].
\end{align}
The three terms in the numerator always sum to $(\partial_\Theta p^{\mathbf{i}}(t_2))^2$, regardless of $t_1$, so the state at $t_2$ alone fixes the total sensitivity. The decomposition therefore shows not how much sensitivity is available but how it is distributed between the state and process. At $t_1=0$ the decomposition reduces to Sec.~\ref{sec:FIglobal}. The generating state then serves as $\hat{\rho}(t_1)$, so $\partial_\Theta p^{\mathbf{i}}(0)=0$ and all sensitivity resides in the process contribution. For generic $t_1$ both contributions are present.

\begin{figure*}[tbp]
    \centering
    \includegraphics[width=\textwidth]{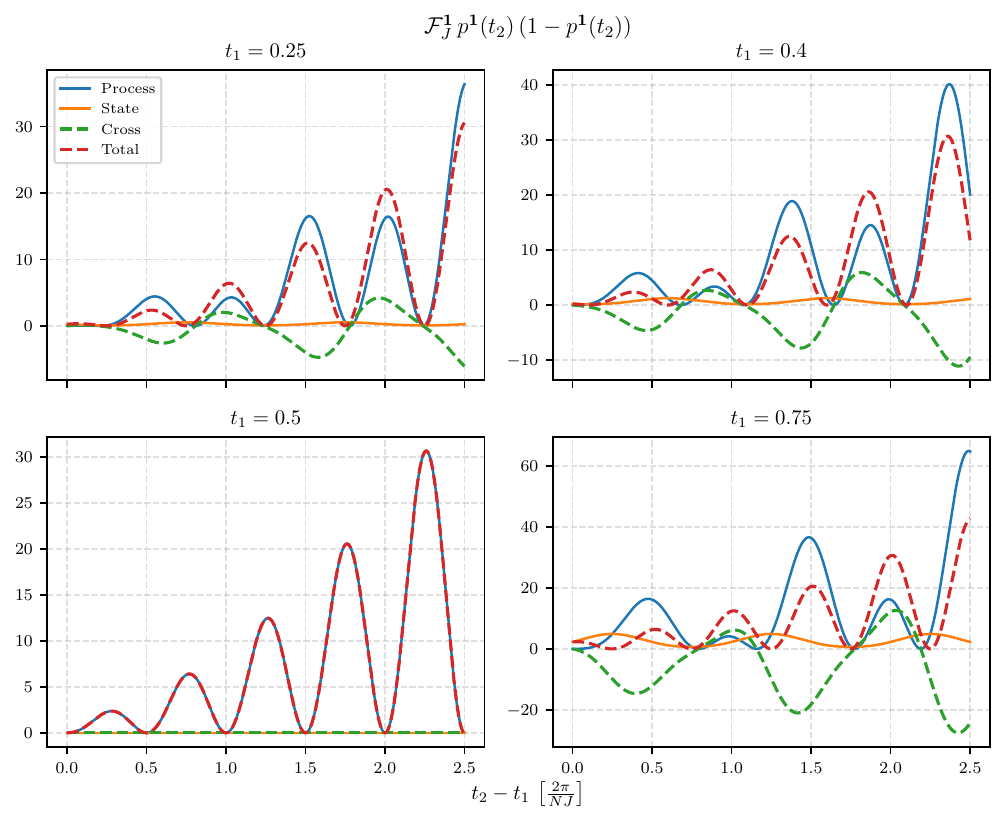}
    \caption{Process, state, cross, and total contributions to the rescaled Fisher information $\mathcal{F}_J^{\mathbf{1}}\,p^{\mathbf{1}}(t_2)(1-p^{\mathbf{1}}(t_2))$ for class~$\mathbf{1}$, plotted against $t_2-t_1$ in units of $2\pi/(NJ)$ for $t_1\in\{0.25, 0.4, 0.5, 0.75\}$, $K=1$, and $N=5$. The cross and total contributions are shown dashed. The total equals $(\partial_J p^{\mathbf{1}}(t_2))^2$ independently of $t_1$.}
    \label{fig:process_state_combined}
\end{figure*}

Figure~\ref{fig:process_state_combined} illustrates the decomposition of the rescaled Fisher information $\mathcal{F}_J^{\mathbf{1}}\,p^{\mathbf{1}}(t_2)(1-p^{\mathbf{1}}(t_2))$ for class~$\mathbf{1}$ at $K=1$, with one panel per value of $t_1$. We plot this quantity rather than $\mathcal{F}_J^{\mathbf{1}}$ itself because the individual process, state, and cross contributions diverge at integer multiples of the period where $p^{\mathbf{1}} \to 1$, while their sum stays finite. Multiplying through by the denominator renders all three contributions finite and directly comparable.

The dashed total is consistent with Fig.~\ref{fig:Fisher_total_N}. At $t_2 = t_1$ the process contribution vanishes and all sensitivity resides in the state contribution, equal to $(\partial_J p^{\mathbf{1}}(t_1))^2$. As $t_2 - t_1$ grows, the process contribution grows secularly and eventually dominates. Class~$\mathbf{0}$ has the same qualitative structure.

The state contribution,
\begin{equation}
\left(1-\varphi_\tau^{\mathbf{i}}\right)^2\!\left(\partial_J p^{\mathbf{i}}(t_1)\right)^2
= \left(\frac{p^{\mathbf{i}}(t_2)}{p^{\mathbf{i}}(t_1)}\right)^2\!\left(\partial_J p^{\mathbf{i}}(t_1)\right)^2,
\end{equation}
is a product of two factors. The factor $(\partial_J p^{\mathbf{i}}(t_1))^2$ is the sensitivity of the state at $t_1$ to $J$, vanishing when $t_1$ is at an integer or half-integer multiple of the period and nonzero otherwise. The factor $(p^{\mathbf{i}}(t_2)/p^{\mathbf{i}}(t_1))^2$ measures how much of that initial sensitivity survives to $t_2$. When the state moves toward the fixed point, $p^{\mathbf{i}}(t_2) < p^{\mathbf{i}}(t_1)$ and the rescaling dilutes the initial sensitivity. When the state moves away from the fixed point, the rescaling amplifies the initial sensitivity. The $t_2$ dependence of the state contribution therefore reflects how the sensitivity already present at $t_1$ is carried forward, not the generation of new information. Since $p^{\mathbf{i}}(t_2)$ is bounded, the state contribution oscillates without secular growth. A smaller $p^{\mathbf{i}}(t_1)$ allows greater amplification when excitation returns, so the state contribution can oscillate more strongly for $t_1=0.4$ than for $t_1=0.25$ despite starting smaller at $t_2=t_1$.

The process contribution,
\begin{equation}
\left(\partial_J\varphi_\tau^{\mathbf{i}}\right)^2\!\left(p^{\mathbf{i}}(t_1)\right)^2
= \left(\partial_J p^{\mathbf{i}}(t_2) - \frac{p^{\mathbf{i}}(t_2)}{p^{\mathbf{i}}(t_1)}\,\partial_J p^{\mathbf{i}}(t_1)\right)^2,
\end{equation}
is the square of the difference between the total sensitivity at $t_2$ and the state sensitivity at $t_1$ carried forward by the rescaling factor. The process contribution measures the sensitivity that is new at $t_2$, beyond what was already encoded at $t_1$. The $t_2$ dependence enters through $\partial_J p^{\mathbf{i}}(t_2) \propto t_2\sin(NJt_2)$, a factor that grows secularly and causes the process contribution to dominate at long times. The process contribution depends on $t_1$ only through the subtracted term $(p^{\mathbf{i}}(t_2)/p^{\mathbf{i}}(t_1))\,\partial_J p^{\mathbf{i}}(t_1)$, a term that stays finite as $t_2$ grows. As $t_2 - t_1$ grows, the process contribution is increasingly dominated by $(\partial_J p^{\mathbf{i}}(t_2))^2$, independent of $t_1$.

The cross contribution is twice the product of the signed process factor $-\partial_\Theta\varphi_\tau^{\mathbf{i}}\,p^{\mathbf{i}}(t_1)$ and the signed state factor $\left(1-\varphi_\tau^{\mathbf{i}}\right)\partial_\Theta p^{\mathbf{i}}(t_1)$, whose squares are the process and state contributions. The cross contribution is positive when the two factors share a sign, so the total exceeds the sum of the process and state contributions, and negative when their signs differ, so the total falls short of that sum. The sign of the cross contribution therefore indicates whether the state at $t_1$ and the dynamics over $[t_1,t_2]$ reinforce or oppose each other in sensing $\Theta$.

The total contribution $(\partial_J p^{\mathbf{i}}(t_2))^2$ vanishes at both integer and half-integer multiples of the period, but $\mathcal{F}_J^{\mathbf{i}}$ itself behaves differently at the two zeros. At half-integer multiples the denominator $p^{\mathbf{i}}(1-p^{\mathbf{i}})$ remains nonzero, so $\mathcal{F}_J^{\mathbf{i}} = 0$, while at integer multiples the numerator and denominator both vanish and the ratio tends to a finite value that grows secularly, as Figs.~\ref{fig:Fisher_total_N} and~\ref{fig:process_state_combined} show. The dips in $\mathcal{F}_\Theta$ at half-integer multiples of the period align with the regime where all future propagators are nonpositive and non-CP, while the maxima occur near the integer multiples where all propagators are positive and CP, shifted slightly forward by the secular growth.

Non-Markovian dynamics and Fisher information are related in ways that depend on how the parameter enters. When the parameter of interest enters only through the initial state of the subsystem, non-Markovianity allows more information to be extracted at late times~\cite{Lu:2010}, since Fisher information is nonincreasing under CPTP maps~\cite{Akio:2001}. Here $J$ governs the global Hamiltonian and therefore enters both through the state at $t_1$ and through the open dynamics over $[t_1,t_2]$. We find no systematic correlation between violations of positivity or the entanglement entropy and the quantum Fisher information, consistent with prior work on non-Markovian open systems~\cite{Mirkin:2020}. Our propagator framing is complementary to work on Fisher information for parameters of open-system master equations~\cite{Vatasescu:2022}. 

\section{Conclusion}\label{sec:conclusion}

Using an explicit $N$-qubit model with a single conserved excitation, and excluding excitation flow in sectors of two or more excitations, we derived closed-form propagators for every $K$-qubit subsystem over every time window. The single transition amplitude $u_d(t)$ governs all open dynamics, so excitation flow, positivity and complete positivity, entanglement entropy, and quantum Fisher information are all functions of one time-dependent quantity.

The trace distance criterion [Eq.~\eqref{eq:pop_criterion}] establishes that a propagator is positive and CP if and only if the trace distance to the fixed point does not increase over $[t_1,t_2]$. Neither the entanglement entropy between the subsystem and the rest of the network at $t_1$ nor its change over $[t_1,t_2]$ has any direct correspondence to that criterion. Positivity and complete positivity coincide for all subsystem sizes and both dynamical classes. The coincidence rests on $\hat{\phi}_\tau^{\mathbf{i}}$ having rank one, and Sec.~\ref{sec:pcp} gives the conditions that keep it there. Nonmonotonicity of the trace distance should persist regardless, since a closed finite network returns excitation to the excited qubit whatever the coupling profile.

The ensemble of propagators constrains global properties that no single subsystem can access. Within each dynamical class, all propagators share one fixed point, and at $K>1$ this fixed point lies in a larger fixed manifold. An observer who finds this common fixed point across several propagators can infer that the global Hamiltonian is symmetric. A band of states lies inside the domain of positivity of every single-qubit propagator, yet the reduced state never visits these states. An ensemble of positive- and CP-divisible dynamics has no such band, since every propagator there preserves positivity on the full state space.

The quantum Fisher information for global parameters decomposes into a classical contribution from the eigenvalues and a quantum contribution from the eigenvectors, with the two sharing the $t^2$ secular growth. Over a time window $[t_1,t_2]$ the Fisher information decomposes into state, process, and cross contributions, with the state contribution bounded while the process contribution grows secularly. The total Fisher information is minimal when all future propagators are nonpositive and non-CP, and near its maximum when they are positive and CP, with secular growth shifting the maxima slightly forward. That alignment is specific to the Hamiltonian, generating state, and single-excitation restriction considered here, though the state--process decomposition holds generally, with different subsystems of a less homogeneous network carrying different Fisher information. Beyond the alignment of the Fisher information minima with the nonpositive regime, we find no systematic relationship between the positivity of the propagators and the sensitivity to global parameters.

This example serves as a benchmark for developing an open-system framework for observers and inference. Such a framework must embrace non-Markovian dynamics, including nonpositive and non-CP propagators, and abandon privileged initial times, reference states, and fixed subsystem boundaries. Both the dynamical prescription and the inference questions differ from their laboratory counterparts, where the experimenter controls either the probe or the dynamics. The broader question is whether we can reformulate open dynamics and inference not in terms of control but in terms of consistency constraints on ensembles, where the ensemble defines the space of global quantum systems compatible with the partial observations available to an observer.

\begin{acknowledgments}
We are grateful for conversations with Zhen Bi, Eugenio Bianchi, Stefan Eccles, and Xiantao Li at Penn State, and with Alexander Rothkopf at the Mainz Institute for Theoretical Physics of the Cluster of Excellence PRISMA+ (Project ID 390831469). This work was supported by the National Science Foundation under Grant No. NSF-PHY-2310662.
\end{acknowledgments}

\appendix

\section{\texorpdfstring{Diagonalizing the\\Single-Excitation Hamiltonian}{Diagonalizing the Single-Excitation Hamiltonian}}\label{app:single}

This appendix derives $u_s(t)$ and $u_d(t)$ of Eq.~\eqref{eq:us_ud} by diagonalizing the Hamiltonian in the single-excitation sector. Excitation conservation restricts Eq.~\eqref{eq:XXZ_hamiltonian} to the $N\times N$ matrix
\begin{equation}
\hat{H}^{(q=1)}=\sum_{i,j=1}^N H_{ij}\ket{i}\bra{j},
\end{equation}
where $\ket{i}$ is the single-excitation state with qubit $i$ excited and every other qubit unexcited. The diagonal entries $H_{ii}$ come from the $\hat{\sigma}^z_i\hat{\sigma}^z_j$ and $\hat{\sigma}^z_i$ terms of Eq.~\eqref{eq:XXZ_hamiltonian}, since $\hat{\sigma}^z$ is diagonal. On $\ket{i}$ the excited qubit carries $\hat{\sigma}^z$ eigenvalue $-1$ and each of the $N-1$ unexcited qubits carries $+1$, so
\begin{equation}
     h\sum_{j=1}^N\hat{\sigma}^z_j\ket{i}=h\left[-1+(N-1)\right]\ket{i}=h(N-2)\ket{i}.
\end{equation}
Among the ordered pairs of $\sum_{i\neq j}$, the $(N-1)(N-2)$ aligned pairs of unexcited qubits each contribute $+1$, while the $2(N-1)$ anti-aligned pairs of the excited qubit with an unexcited qubit each contribute $-1$, so the $\hat{\sigma}^z_i\hat{\sigma}^z_j$ term contributes $J_z(N-1)(N-4)/4$ and
\begin{equation}
H_s = \frac{J_z}{4}(N-1)(N-4)+h(N-2),
\end{equation}
common to every $H_{ii}$. The $\hat{\sigma}^x_i\hat{\sigma}^x_j+\hat{\sigma}^y_i\hat{\sigma}^y_j$ term moves the excitation between qubits $i$ and $j$, and the factor $1/4$ of Eq.~\eqref{eq:XXZ_hamiltonian} together with the two orderings of each pair gives every off-diagonal entry the same value $H_{ij}=J$, independent of $N$ since only these two qubits are involved. Collecting these entries gives
\begin{equation}
\hat{H}^{(q=1)}=\begin{pmatrix}
    H_s & J & \cdots & J\\
    J & H_s & \cdots & J\\
    \vdots & \vdots & \ddots & \vdots\\
    J & J & \cdots & H_s
\end{pmatrix}.
\label{eq:Hq1}
\end{equation}

To exponentiate the Hamiltonian, it is convenient to diagonalize. Acting with Eq.~\eqref{eq:Hq1} on the uniform superposition
\begin{equation}
\ket{+}=\frac{1}{\sqrt{N}}\sum_{i=1}^N\ket{i}
\label{eq:plus_state}
\end{equation}
returns $\left[H_s+(N-1)J\right]\ket{+}$, since every row of Eq.~\eqref{eq:Hq1} carries one $H_s$ and $N-1$ copies of $J$, so $\ket{+}$ is an eigenvector with eigenvalue $H_s+(N-1)J$. Acting on a state $\sum_i c_i\ket{i}$ whose amplitudes satisfy $\sum_i c_i=0$ scales every amplitude by a common factor, $H_sc_i+J\sum_{j\neq i}c_j=(H_s-J)c_i$. The $N-1$ independent states satisfying $\sum_i c_i=0$ span the subspace orthogonal to $\ket{+}$, and each is an eigenvector with eigenvalue $H_s-J$. Those two eigenspaces, of dimension one and $N-1$, exhaust the sector, and their projectors are
\begin{equation}
\hat{P}_+=\ket{+}\bra{+}, \qquad
\hat{P}_-=\hat{\mathbbm{1}}-\hat{P}_+,
\label{eq:projectors}
\end{equation}
and spectral decomposition
\begin{equation}
\hat{H}^{(q=1)}=\left[H_s+(N-1)J\right]\hat{P}_+ + \left(H_s-J\right)\hat{P}_-.
\label{eq:Hq1_spec}
\end{equation}
Exponentiating Eq.~\eqref{eq:Hq1_spec} and factoring out an overall phase gives
\begin{equation}
\hat{U}^{(q=1)}(t)=e^{-i\left[H_s+(N-1)J\right]t}\left[\hat{P}_++e^{iNJt}\hat{P}_-\right],
\label{eq:Uq1_proj}
\end{equation}
where the relative angular frequency $NJ$ is the gap between the two eigenvalues $[H_s+(N-1)J]-(H_s-J)$.

The entries of the two projectors in the excitation basis are
\begin{equation}
\left(\hat{P}_+\right)_{ij}=\frac{1}{N}, \qquad
\left(\hat{P}_-\right)_{ij}=\delta_{ij}-\frac{1}{N}.
\label{eq:proj_entries}
\end{equation}
Writing Eq.~\eqref{eq:Uq1_proj} out entry by entry therefore gives the amplitude for the excitation to remain on the same qubit along the diagonal and the transition amplitude between different qubits off the diagonal, both named in Sec.~\ref{sec:nqubit},
\begin{equation}
\hat{U}^{(q=1)}(t)=e^{-i\left[H_s+(N-1)J\right]t}\begin{pmatrix}
    u_s & u_d & \cdots & u_d\\
    u_d & u_s & \cdots & u_d\\
    \vdots & \vdots & \ddots & \vdots\\
    u_d & u_d & \cdots & u_s
\end{pmatrix},
\label{eq:Uq1}
\end{equation}
with
\begin{equation}
u_s=\frac{1+(N-1)e^{iNJt}}{N}, \qquad
u_d=\frac{1-e^{iNJt}}{N}.
\tag*{(\ref{eq:us_ud})}
\end{equation}

The factor $N-1$ multiplying $e^{iNJt}$ in $u_s$ is the dimension of the degenerate eigenspace, and the minus sign in $u_d$ carries the sign of the off-diagonal entries of $\hat{P}_-$. The overall phase $e^{-i[H_s+(N-1)J]t}$ cancels in $\hat{\rho}(t)=\hat{U}(t)\,\hat{\rho}(0)\,\hat{U}^\dagger(t)$ for any value of $H_s$, so $u_s$ and $u_d$ in Eq.~\eqref{eq:us_ud} depend only on $N$, $J$, and $t$.

\section{Vectorized Form of the Dynamical Map}\label{app:maps}

This appendix derives the vectorized form of the dynamical map $\Phi(0,t)$ of Eq.~\eqref{eq:reduced_map}. Appendix~\ref{app:prop} composes this map via Eq.~\eqref{eq:compose} into the propagator $\Phi(t_1,t_2)$.

The generating state is a product state of the subsystem $S$ and the rest of the network $E$, and we denote the state of $E$ by $\ket{r_0}$. Tracing over $E$ in a basis $\{\ket{r}\}$ then puts the open dynamics in Kraus form,
\begin{equation}\label{eq:kraus}
\begin{split}
\hat{\rho}_S(t)&=\operatorname{Tr}_E\!\left[\hat{U}(t)\,\hat{\rho}_S(0)\otimes\ket{r_0}\bra{r_0}\,\hat{U}^\dagger(t)\right]\\
&=\sum_r\hat{V}_r(t)\,\hat{\rho}_S(0)\,\hat{V}_r^\dagger(t),
\end{split}
\end{equation}
where
\begin{equation}\label{eq:Vr_def}
\hat{V}_r(t)=\bra{r}\hat{U}(t)\ket{r_0}
\end{equation}
acts on the Hilbert space of $S$. Writing out the matrix elements of Eq.~\eqref{eq:kraus} gives
\begin{equation}\label{eq:matelem}
\left(\hat{\rho}_S(t)\right)_{\alpha\beta}
=\sum_r\sum_{\mu,\nu}(\hat{V}_r)_{\alpha\mu}(\hat{V}_r^*)_{\beta\nu}\left(\hat{\rho}_S(0)\right)_{\mu\nu}.
\end{equation}

Stacking the rows of $\hat{\rho}_S$ into a column vector,
\begin{equation}
\vec{\rho}_S=
\begin{pmatrix}
    \rho_{00}\\
    \rho_{01}\\
    \vdots\\
    \rho_{10}\\
    \vdots
\end{pmatrix},
\end{equation}
turns Eq.~\eqref{eq:matelem} into a matrix acting on $\vec{\rho}_S$. We write $V_{\alpha\mu}\equiv(\hat{V}_r)_{\alpha\mu}$ as shorthand. At $K=1$, where $\alpha$ and $\mu$ label the single-qubit basis of $S$, that matrix is
\begin{equation}
\vec{\rho}_S(t)=
\sum_r
{\begin{pmatrix}
V_{00}V^*_{00} & V_{00}V^*_{01} & V_{01}V^*_{00} & V_{01}V^*_{01}\\
V_{00}V^*_{10} & V_{00}V^*_{11} & V_{01}V^*_{10} & V_{01}V^*_{11}\\
V_{10}V^*_{00} & V_{10}V^*_{01} & V_{11}V^*_{00} & V_{11}V^*_{01}\\
V_{10}V^*_{10} & V_{10}V^*_{11} & V_{11}V^*_{10} & V_{11}V^*_{11}
\end{pmatrix}}_r
\vec{\rho}_S(0),
\end{equation}
and grouping the entries by their unstarred index gives the Kronecker product,
\begin{equation}
{\begin{pmatrix}
V_{00}\begin{pmatrix}
V^*_{00} & V^*_{01}\\
V^*_{10} & V^*_{11}
\end{pmatrix} & V_{01}\begin{pmatrix}
V^*_{00} & V^*_{01}\\
V^*_{10} & V^*_{11}
\end{pmatrix}\\
V_{10}\begin{pmatrix}
V^*_{00} & V^*_{01}\\
V^*_{10} & V^*_{11}
\end{pmatrix} & V_{11}\begin{pmatrix}
V^*_{00} & V^*_{01}\\
V^*_{10} & V^*_{11}
\end{pmatrix}
\end{pmatrix}}_r
=\hat{V}_r\otimes\hat{V}_r^*.
\end{equation}
This identification holds at any $K$. The matrix element of Eq.~\eqref{eq:matelem} is then the Kronecker entry
\begin{equation}
\left(\hat{V}_r\otimes\hat{V}_r^*\right)_{(\alpha\beta),(\mu\nu)}
=(\hat{V}_r)_{\alpha\mu}(\hat{V}_r^*)_{\beta\nu}.
\end{equation}
For a $K$-qubit subsystem, the dynamical map is therefore the $4^K\times4^K$ matrix
\begin{equation}
\begin{gathered}
    \vec{\rho}_S(t) = \hat{\Phi}^{\mathbf{i}}(0,t)\,\vec{\rho}_S(0), \\
    \hat{\Phi}^{\mathbf{i}}(0,t) = \sum_r \hat{V}_r(t)\otimes\hat{V}_r^*(t).
\end{gathered}
\end{equation}

Here the generating state fixes $\ket{r_0}$ separately for each dynamical class. For class~$\mathbf{1}$ the excitation lies entirely inside $S$, leaving $E$ in the ground state $\ket{\mathbf{0}}$. For class~$\mathbf{0}$ the excitation lies on a qubit $i$ of $E$, giving $\ket{\mathbf{1}_i}$,
\begin{equation}\label{eq:r0}
\ket{r_0}=\begin{cases}
\ket{\mathbf{0}}, & \text{class }\mathbf{1},\\
\ket{\mathbf{1}_i}, & \text{class }\mathbf{0}.
\end{cases}
\end{equation}

For class~$\mathbf{1}$, excitation conservation allows only states $\ket{r}$ of excitation number at most one, so the sum over $\ket{r}$ in Eq.~\eqref{eq:kraus} runs over $\ket{\mathbf{0}}$ and the $N-K$ states $\ket{\mathbf{1}_i}$ with $i$ a qubit of $E$. Here $\ket{r_0}=\ket{\mathbf{0}}$, so $\hat{V}^{\mathbf{1}}=\bra{\mathbf{0}}\hat{U}\ket{\mathbf{0}}$ leaves $E$ in the ground state, and $\bra{\mathbf{1}_i}\hat{U}\ket{\mathbf{0}}$ carries the excitation out of $S$. Permutation symmetry makes these $N-K$ operators equal to a single operator $\hat{V}_\tau^{\mathbf{1}}=\bra{\mathbf{1}_i}\hat{U}\ket{\mathbf{0}}$,
\begin{equation}
\hat{\Phi}^{\mathbf{1}}(0,t)=\hat{V}^{\mathbf{1}}\otimes\hat{V}^{\mathbf{1}*}
+(N-K)\,\hat{V}_\tau^{\mathbf{1}}\otimes\hat{V}_\tau^{\mathbf{1}*}.
\end{equation}
At $K=1$, with the first slot of each bra and ket labeling $S$ and the second labeling $E$, these are the $2\times2$ operators
\begin{equation}
\begin{gathered}
\hat{V}^{\mathbf{1}}=\begin{pmatrix}
\bra{0\,\mathbf{0}}\hat{U}\ket{0\,\mathbf{0}} & \bra{0\,\mathbf{0}}\hat{U}\ket{1\,\mathbf{0}}\\
\bra{1\,\mathbf{0}}\hat{U}\ket{0\,\mathbf{0}} & \bra{1\,\mathbf{0}}\hat{U}\ket{1\,\mathbf{0}}
\end{pmatrix}
=\begin{pmatrix}u_0&\\&u_s\end{pmatrix},\\
\hat{V}_\tau^{\mathbf{1}}=\begin{pmatrix}
\bra{0\,\mathbf{1}_i}\hat{U}\ket{0\,\mathbf{0}} & \bra{0\,\mathbf{1}_i}\hat{U}\ket{1\,\mathbf{0}}\\
\bra{1\,\mathbf{1}_i}\hat{U}\ket{0\,\mathbf{0}} & \bra{1\,\mathbf{1}_i}\hat{U}\ket{1\,\mathbf{0}}
\end{pmatrix}
=\begin{pmatrix}&u_d\\\phantom{u_d}&\end{pmatrix}.
\end{gathered}
\end{equation}
Assembling the tensor products gives
\begin{equation}
\hat{\Phi}^{\mathbf{1}}(0,t)=\begin{pmatrix}
1 & & & (N-1)\left|u_d\right|^2\\
& u_0u_s^* & &\\
& & u_0^*u_s &\\
& & & \left|u_s\right|^2
\end{pmatrix}.
\end{equation}

For class~$\mathbf{0}$, excitation conservation in principle allows $\ket{r}$ of higher excitation number, since the input state of $S$ can carry excitations of its own. Under the restriction of Sec.~\ref{sec:singleChargeSector}, however, every sector of excitation number above one evolves by a phase alone, removing the off-diagonal amplitudes such as $u_{2d}$ and leaving the sum in Eq.~\eqref{eq:kraus} running over excitation number at most one. Here $\ket{r_0}=\ket{\mathbf{1}_i}$ for qubit $i$ of $E$, so the sum over $\ket{r}$ runs over $\ket{\mathbf{1}_i}$ itself, the $N-K-1$ states $\ket{\mathbf{1}_j}$ on the other qubits $j\neq i$ of $E$, and $\ket{\mathbf{0}}$. This gives $\hat{V}^{\mathbf{0}}=\bra{\mathbf{1}_i}\hat{U}\ket{\mathbf{1}_i}$, leaving the excitation on qubit $i$, $\bra{\mathbf{1}_j}\hat{U}\ket{\mathbf{1}_i}$, moving the excitation to another qubit of $E$, and $\hat{V}_\tau^{\mathbf{0}}=\bra{\mathbf{0}}\hat{U}\ket{\mathbf{1}_i}$, taking the excitation into $S$. Permutation symmetry makes the $N-K-1$ operators equal to a single operator $\hat{V}_0^{\mathbf{0}}=\bra{\mathbf{1}_j}\hat{U}\ket{\mathbf{1}_i}$,
\begin{equation}
\hat{\Phi}^{\mathbf{0}}(0,t)=\hat{V}^{\mathbf{0}}\otimes\hat{V}^{\mathbf{0}*}
+(N-K-1)\,\hat{V}_0^{\mathbf{0}}\otimes\hat{V}_0^{\mathbf{0}*}
+\hat{V}_\tau^{\mathbf{0}}\otimes\hat{V}_\tau^{\mathbf{0}*}.
\end{equation}
At $K=1$, these are
\begin{equation}
\begin{gathered}
\hat{V}^{\mathbf{0}}=\begin{pmatrix}
\bra{0\,\mathbf{1}_i}\hat{U}\ket{0\,\mathbf{1}_i} & \bra{0\,\mathbf{1}_i}\hat{U}\ket{1\,\mathbf{1}_i}\\
\bra{1\,\mathbf{1}_i}\hat{U}\ket{0\,\mathbf{1}_i} & \bra{1\,\mathbf{1}_i}\hat{U}\ket{1\,\mathbf{1}_i}
\end{pmatrix}
=\begin{pmatrix}u_s&\\&u_2\end{pmatrix},\\
\hat{V}_0^{\mathbf{0}}=\begin{pmatrix}
\bra{0\,\mathbf{1}_j}\hat{U}\ket{0\,\mathbf{1}_i} & \bra{0\,\mathbf{1}_j}\hat{U}\ket{1\,\mathbf{1}_i}\\
\bra{1\,\mathbf{1}_j}\hat{U}\ket{0\,\mathbf{1}_i} & \bra{1\,\mathbf{1}_j}\hat{U}\ket{1\,\mathbf{1}_i}
\end{pmatrix}
=\begin{pmatrix}u_d&\\&\phantom{u_d}\end{pmatrix},\\
\hat{V}_\tau^{\mathbf{0}}=\begin{pmatrix}
\bra{0\,\mathbf{0}}\hat{U}\ket{0\,\mathbf{1}_i} & \bra{0\,\mathbf{0}}\hat{U}\ket{1\,\mathbf{1}_i}\\
\bra{1\,\mathbf{0}}\hat{U}\ket{0\,\mathbf{1}_i} & \bra{1\,\mathbf{0}}\hat{U}\ket{1\,\mathbf{1}_i}
\end{pmatrix}
=\begin{pmatrix}&\phantom{u_d}\\u_d&\end{pmatrix}.
\end{gathered}
\end{equation}
Assembling gives
\begin{equation}
\hat{\Phi}^{\mathbf{0}}(0,t)=\begin{pmatrix}
1-\left|u_d\right|^2 & & &\\
& u_su_2^* & &\\
& & u_s^*u_2 &\\
\left|u_d\right|^2 & & & 1
\end{pmatrix}.
\end{equation}
In both classes $\hat{V}_\tau^{\mathbf{i}}$ carries the excitation across the subsystem boundary, out for class~$\mathbf{1}$ and in for class~$\mathbf{0}$.

\section{\texorpdfstring{Propagator Composition\\and Trace Preservation}{Propagator Composition and Trace Preservation}}\label{app:prop}

In the vectorized form of Appendix~\ref{app:maps}, the composition of Eq.~\eqref{eq:compose} becomes matrix multiplication,
\begin{equation}\label{eq:compose_vec}
\hat{\Phi}^{\mathbf{i}}(t_1,t_2)=\hat{\Phi}^{\mathbf{i}}(0,t_2)\,\hat{\Phi}^{\mathbf{i}}(0,t_1)^{-1}.
\end{equation}

The inverse in Eq.~\eqref{eq:compose_vec} exists at all but isolated points. The sole exceptions occur when the subsystem and the rest of the network have equal size and the excitation is maximally delocalized, at $K=N/2$ and $t_1=(m+1/2)\times 2\pi/(NJ)$ with $m\in\mathbb{Z}$, where a denominator in the elements of $\Phi(0,t_1)$ vanishes. The singular behavior is well-defined in terms of left and right limits and does not affect any result.

At $K=1$, Eq.~\eqref{eq:compose_vec} gives
\begin{equation}
\begin{gathered}
\hat{\Phi}^{\mathbf{1}}(t_1,t_2)=
\begin{pmatrix}
    1 & & & \varphi_\tau^{\mathbf{1}}\\
    & \varphi_0^{\mathbf{1}}\varphi_s^{\mathbf{1}*} & &\\
    & & \varphi_0^{\mathbf{1}*}\varphi_s^{\mathbf{1}} &\\
    & & & \left|\varphi_s^{\mathbf{1}}\right|^2
\end{pmatrix},\\
\varphi_s^{\mathbf{1}}=\frac{u_s(t_2)}{u_s(t_1)}, \qquad
\varphi_\tau^{\mathbf{1}}=1-\left|\varphi_s^{\mathbf{1}}\right|^2,\\
\left|\varphi_0^{\mathbf{1}}\right|=1,
\end{gathered}
\end{equation}
and
\begin{equation}
\begin{gathered}
\hat{\Phi}^{\mathbf{0}}(t_1,t_2)=
\begin{pmatrix}
    \varphi_0^{\mathbf{0}}& & &\\
    &\varphi_s^{\mathbf{0}}\varphi_2^{\mathbf{0}*}& &\\
    & &\varphi_s^{\mathbf{0}*}\varphi_2^{\mathbf{0}}&\\
    \varphi_\tau^{\mathbf{0}}& & &1
\end{pmatrix},\\
\varphi_0^{\mathbf{0}}=\frac{1-|u_d(t_2)|^2}{1-|u_d(t_1)|^2}, \qquad
\varphi_\tau^{\mathbf{0}}=1-\varphi_0^{\mathbf{0}},\\
\varphi_s^{\mathbf{0}}=\frac{u_s(t_2)}{u_s(t_1)}, \qquad 
\left|\varphi_2^{\mathbf{0}}\right|=1.
\end{gathered}
\end{equation}
Here $\hat{\Phi}^{\mathbf{1}}(t_1,t_2)$ and $\hat{\Phi}^{\mathbf{0}}(t_1,t_2)$ are the single-qubit propagators written in the excitation basis, while $\hat{\Lambda}^{\mathbf{1}}(t_1,t_2)$ and $\hat{\Lambda}^{\mathbf{0}}(t_1,t_2)$ of Sec.~\ref{sec:geometry} are written in the Bloch-vector representation.

For a $K$-qubit subsystem, Eq.~\eqref{eq:compose_vec} composes the propagator entries
\begin{equation}
\begin{gathered}
\varphi_s^{\mathbf{1}}=\frac{u_d\!\left(t_1\right)u_d\!\left(t_2\right)-u_s\!\left(t_1\right)u_s\!\left(t_2\right)}{\big[u_d\!\left(t_1\right)-u_s\!\left(t_1\right)\big]\big[(K-1)u_d\!\left(t_1\right)+u_s\!\left(t_1\right)\big]}\\
\hspace{2em}+\frac{(K-2)u_d\!\left(t_1\right)\big[u_d\!\left(t_2\right)-u_s\!\left(t_2\right)\big]}{\big[u_d\!\left(t_1\right)-u_s\!\left(t_1\right)\big]\big[(K-1)u_d\!\left(t_1\right)+u_s\!\left(t_1\right)\big]},\\
\varphi_d^{\mathbf{1}}=\frac{u_d\!\left(t_1\right)u_s\!\left(t_2\right)-u_s\!\left(t_1\right)u_d\!\left(t_2\right)}{\big[u_d\!\left(t_1\right)-u_s\!\left(t_1\right)\big]\big[(K-1)u_d\!\left(t_1\right)+u_s\!\left(t_1\right)\big]},\\
\left|\varphi_q^{\mathbf{1}}\right|=1, \qquad q\in\{0,2,3,\ldots,K\},
\end{gathered}
\tag*{(\ref{eq:phi_sdq1})}
\end{equation}
\begin{equation}
\begin{gathered}
\varphi_s^{\mathbf{0}}=\frac{u_s\!\left(t_2\right)}{u_s\!\left(t_1\right)},\\
\left|\varphi_{q+1}^{\mathbf{0}}\right|=1,
\qquad q\in\{1,2,\ldots,K\},
\end{gathered}
\tag*{(\ref{eq:phi_sq0})}
\end{equation}
\begin{equation}
\varphi_0^{\mathbf{0}}=\frac{1-K\left|u_d\!\left(t_2\right)\right|^2}{1-K\left|u_d\!\left(t_1\right)\right|^2},
\tag*{(\ref{eq:phi_00})}
\end{equation}
together with the excitation-flow parameters
\begin{equation}
\varphi_\tau^{\mathbf{1}}
= \frac{\left|u_d\!\left(t_2\right)\right|^2-\left|u_d\!\left(t_1\right)\right|^2}
       {\frac{1}{N-K}-K\!\left|u_d\!\left(t_1\right)\right|^2},
\tag*{(\ref{eq:phi_tau1})}
\end{equation}
and
\begin{equation}
\varphi_\tau^{\mathbf{0}}=\frac{\left|u_d\!\left(t_2\right)\right|^2-\left|u_d\!\left(t_1\right)\right|^2}{1-K\left|u_d\!\left(t_1\right)\right|^2}.
\tag*{(\ref{eq:phi_tau0})}
\end{equation}

Inverting $\hat{\Phi}^{\mathbf{i}}(0,t_1)$ accounts for the two forms these expressions take. An entry alone on the diagonal inverts to its reciprocal, so the composition divides one time by the other and returns the ratios $\varphi_s^{\mathbf{0}}$ and $\varphi_0^{\mathbf{0}}$. Entries away from the diagonal enter the inverse through minors of alternating sign and therefore compose into differences. In the $q=1$ sector of class~$\mathbf{1}$, $\varphi_d^{\mathbf{1}}$ redistributes the excitation among the $K$ qubits, so $\varphi_s^{\mathbf{1}}$ and $\varphi_d^{\mathbf{1}}$ carry differences of products of $u_s$ and $u_d$. The entries built from $\hat{V}_\tau^{\mathbf{i}}$ carry excitation across the subsystem boundary and also lie off the diagonal, so the numerator of $\varphi_\tau^{\mathbf{i}}$ is the difference $|u_d(t_2)|^2-|u_d(t_1)|^2$ and sets its sign.

Undoing the vectorization of $\hat{\rho}_S$ from Appendix~\ref{app:maps} converts $\hat{\Phi}^{\mathbf{i}}(t_1,t_2)$ back into the operator-sum form
\begin{equation}
\Phi\!\left(t_1,t_2\right)\left[\hat{\rho}\right]
=\hat{\phi}\,\hat{\rho}\,\hat{\phi}^{\dagger}+\sum_i\hat{\phi}_i\,\hat{\rho}\,\hat{\phi}_i^T
\tag*{(\ref{eq:MSR})}
\end{equation}
established for each class in Sec.~\ref{sec:env-zero} and Sec.~\ref{sec:env-one}. The conversion places one factor of $\sqrt{\varphi_\tau^{\mathbf{i}}}$ on each side of $\hat{\rho}_S$. A dagger would pair them as $|\sqrt{\varphi_\tau^{\mathbf{i}}}|^{2}=|\varphi_\tau^{\mathbf{i}}|$ and lose the sign, so the second term of Eq.~\eqref{eq:MSR} above uses the transpose, returning $(\sqrt{\varphi_\tau^{\mathbf{i}}})^{2}=\varphi_\tau^{\mathbf{i}}$. Class~$\mathbf{0}$'s $\hat{\phi}_0^{\mathbf{0}}$ carries the analogous entry $\sqrt{\varphi_0^{\mathbf{0}}-|\varphi_s^{\mathbf{0}}|^2}$, real or imaginary since $\varphi_0^{\mathbf{0}}-|\varphi_s^{\mathbf{0}}|^2$ can likewise be negative.

The composed propagator must also preserve the trace of $\hat{\rho}$. Requiring
\begin{equation}
\operatorname{Tr}\left[\Phi^{\mathbf{i}}(t_1,t_2;K)\!\left[\hat{\rho}\right]\right]=\operatorname{Tr}\left[\hat{\rho}\right]
\end{equation}
for arbitrary $\hat{\rho}$ gives the trace-preservation conditions
\begin{equation}
\begin{gathered}
|\varphi_s^{\mathbf{1}}|^2 + (K-1)\,|\varphi_d^{\mathbf{1}}|^2 + \varphi_\tau^{\mathbf{1}} = 1,\\
2\Re\!\left[\varphi_s^{\mathbf{1}*} \varphi_d^{\mathbf{1}}\right] + (K-2)|\varphi_d^{\mathbf{1}}|^2 + \varphi_\tau^{\mathbf{1}} = 0,
\end{gathered}
\tag*{(\ref{eq:trace_pres_1})}
\end{equation}
and
\begin{equation}
\varphi_0^{\mathbf{0}} + K\,\varphi_\tau^{\mathbf{0}} = 1.
\tag*{(\ref{eq:trace_pres_0})}
\end{equation}
Substituting the expressions for $u_s$ and $u_d$ of Eq.~\eqref{eq:us_ud} confirms both conditions hold for all $t_1$ and $t_2$, consistent with the unitarity of the global network.

We verified the propagator composition of Eqs.~\eqref{eq:phi_sdq1}, \eqref{eq:phi_sq0}, \eqref{eq:phi_00}, \eqref{eq:phi_tau1}, and \eqref{eq:phi_tau0} explicitly for $K\in\{1,2,3\}$ and extrapolated to arbitrary $K$ rather than proving the pattern for all $K$.

\section{Spectrum of the Choi Matrix}\label{app:choi}

To derive the condition for complete positivity, we diagonalize the Choi matrix $\hat{C}_\Phi$. For class~$\mathbf{1}$ the spectrum is
\begin{equation}
\left\{2^K-K\varphi_\tau^{\mathbf{1}},\ K\varphi_\tau^{\mathbf{1}},\ 0,\ \ldots,\ 0\right\},
\end{equation}
and since Eq.~\eqref{eq:trace_pres_1} bounds $\varphi_\tau^{\mathbf{1}}\leq1$, we have $2^K-K\varphi_\tau^{\mathbf{1}}\geq1$, so only $K\varphi_\tau^{\mathbf{1}}$ can be negative.

For class~$\mathbf{0}$ the spectrum is
\begin{equation}
\begin{gathered}
\left\{\lambda_+,\ \lambda_-,\ K\varphi_\tau^{\mathbf{0}},\ 0,\ \ldots,\ 0\right\}, \\
\lambda_\pm=\frac{1}{2}\left[\left(2^K-1+\varphi_0^{\mathbf{0}}\right)\pm\sqrt{\Delta}\right], \\
\Delta=\left(2^K-1+\varphi_0^{\mathbf{0}}\right)^2-4\left(2^K-1\right)\frac{N-1-K}{\left|u_s(t_1)\right|^2}\varphi_\tau^{\mathbf{0}}.
\end{gathered}
\end{equation}
Since Eq.~\eqref{eq:trace_pres_0} gives $\varphi_0^{\mathbf{0}}=1-K\varphi_\tau^{\mathbf{0}}$, the eigenvalue $\lambda_+$ stays positive, so for $K\leq N-1$ only $K\varphi_\tau^{\mathbf{0}}$ and $\lambda_-$ can be negative, both when and only when $\varphi_\tau^{\mathbf{0}}<0$, with $\lambda_-$ reaching zero at $K=N-1$.

\bibliography{main}

\end{document}